\documentclass[10pt]{iopart}
\usepackage{iopams}  
\usepackage{amssymb}
\usepackage[T1]{fontenc}
\usepackage{setstack}
\usepackage{graphicx}
\usepackage{bm}
\usepackage{color}
\usepackage{hyperref}
\usepackage{braket}
\usepackage{comment}
\usepackage{inputenc}
\usepackage{comment}
\usepackage{bbm}
\usepackage{cite}
\begin{document}
	
	\title[Hyperbolic Plasmon dispersion and Optical Conductivity of Holey Graphene]{Hyperbolic Plasmon dispersion and Optical Conductivity of Holey Graphene: signatures of flat-bands}
	
	\author{Abdiel de Jesús Espinosa-Champo$^{1,2,3}$ and Gerardo G. Naumis$^{3}$}

	\address{${}^{1}$Posgrado de Ciencias F\'isicas, Universidad Nacional Autónoma de México, Apartado Postal 20-364 01000, Ciudad de México, México.}
	\address{${}^{2}$ Departamento de F\'isica, Facultad de Ciencias, Universidad Nacional Aut\'onoma de M\'exico, Apdo. Postal 70-542, 04510, CDMX, México.}
	\address{${}^{3}$Depto. de Sistemas Complejos, Instituto de F\'isica, Universidad Nacional Aut\'onoma de M\'exico (UNAM). Apdo. Postal 20-364, 01000, CDMX, Mexico.}
	\ead{naumis@fisica.unam.mx}
	\vspace{10pt}
	\begin{indented}
		\item[]\today
	\end{indented}
	
	\begin{abstract}
Holey graphene (HG) is a novel two-dimensional (2D) material that has attracted considerable attention due to its remarkable electrical, thermal, and mechanical properties. The recent discovery of flat bands in HG has garnered significant interest. In this work, we systematically investigate the tunable plasmonic modes and optical conductivity of HG at or near the flat band condition by changing the holes radii and periodic configuration. It is found that HG presents nearly flat plasmonic bands in configurations with larger hole radii. Hyperbolic plasmons are found due to the breaking of the graphene's bipartite sublattice symmetry induced by the holes. Such an effect is also confirmed by looking at the optical conductivity, that also presents a marked anisotropy. The material's marked optical anisotropy leads to hyperbolic plasmons, making it a promising platform for nanophotonic applications.
\end{abstract}

	\maketitle

\section{Introduction} \label{sec: Introduction}
Two-dimensional (2D) materials have revolutionized condensed matter physics and materials science, opening up novel opportunities in electronics, photonics, and quantum technologies \cite{Shanmugam2022,ferrari2015science,Naumis_2017, Naumis_2024}. In particular, graphene, a monolayer of carbon atoms arranged in a hexagonal lattice, has emerged as the prototypical 2D material due to its extraordinary electrical, optical, thermal, and mechanical properties \cite{Naumis_2017,Naumis_2024,MBAYACHI2021100163,novoselov2005two,geim2007rise,cao2018unconventional,Abdiel2018,Abdiel2021,Abdiel2024,bonaccorso2010graphene,koppens2014photodetectors, kim2012graphene}. 

One of the most promising aspects of graphene is its unique plasmonic behavior \cite{Hwang2007,Ogawa2020, Principi2018,Yu2016}. Plasmons, the collective oscillations of free charge carriers \cite{Pines1952, Pines1962}, are crucial for applications in nanophotonics, as they enable the confinement of electromagnetic energy to scales smaller than the diffraction limit \cite{Yu2019, Yu2016,Paudel17}. In graphene, plasmons exhibit long lifetimes and high confinement \cite{Principi2018,Yu2016}.

However, the zero band gap of pristine graphene limits its direct application in optoelectronic devices, where a tunable band gap is essential \cite{YU2015103,LUO2022109290}. To overcome this limitation, various modifications of graphene's structure have been pursued, both theoretically, including twisted bilayer graphene \cite{Stauber2016}, twisted double bilayer graphene \cite{Chakraborty2022}, trilayer graphene under pressure \cite{Wu2021}, or multilayer graphene on a conducting substrate \cite{Gumbs2016}, and experimentally, including graphene-covered gold nanovoid arrays \cite{Zhu2013} and holey graphene (HG) \cite{Miao2019,Zhu2014, Liu2015, Nikitin2012}, this designs results in a significantly stronger optical response and allow to host plasmon modes that are highly tunable in both periodicity and size of the holes \cite{Miao2019,Zhu2014}.

Therefore, holey graphene (HG) has attracted particular interest \cite{Lin04072017, Moreno2018,Pedersen20081,Pedersen2008,Mahmood2015, Xu2019}. HG is a variant of graphene in which nanoscale holes are periodically introduced into the lattice \cite{Lin2023}. These intentional vacancies can break symmetries, such as sublattice and inversion symmetries \cite{Champo_2024}. This can significantly alter the electronic, optical, and thermal properties of the material \cite{Singh2020,Pan2023,Pedersen20081}. By tuning the size, shape, and periodicity of these holes, HG can be transformed from a semimetal into a semiconductor with a variable band gap, expanding its potential applications across a range of fields \cite{Pedersen20081, Champo_2024, Barkov2021}. However, a detailed understanding of how the hole geometry precisely dictates the plasmonic dispersion and optical anisotropy, particularly near the flat-band condition, remains an open area of investigation.

Building upon our previous work where we demonstrated the emergence of flat bands in HG \cite{Champo_2024}, here we investigate the direct consequences of these electronic features on the collective excitations and the optical response of the material. This enhances electron-electron interactions and potentially gives rise to novel correlated states \cite{Aoki1993}. Similar phenomena are observed in moir\'e systems such as twisted bilayer graphene (tBLG), where flat bands lead to exotic phases, including superconductivity and correlated insulator states \cite{cao2018unconventional, yankowitz2019tuning}.

This paper is organized as follows,  in Sec. \ref{sec:Hamiltonian Model}, we present the tight-binding model and the unit cells of the HG primarily under discussion. In Sec. \ref{sec:Numerical Methods}, we describe the computational methods used in our study, including the details of the combination of tight-binding propagation method (TBPM) with Kubo formula and Lindhard function to compute the optical conductivities and polarization functions, respectively.  Section \ref{sec:Results and Discussion} presents the results and our analysis, highlighting the effects of different hole patterns on the plasmonic modes and optical conductivity of HG.  Finally, our conclusions and future perspectives are in Sec. \ref{sec:conclusions}.


\section{Hamiltonian Model} \label{sec:Hamiltonian Model}

\begin{figure}
	\centering
	\large{a)}\includegraphics[width=0.4\linewidth]{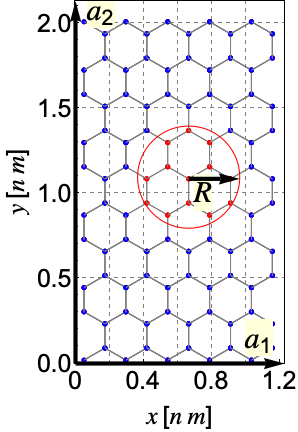}
	\large{b)} \includegraphics[width=0.4\linewidth]{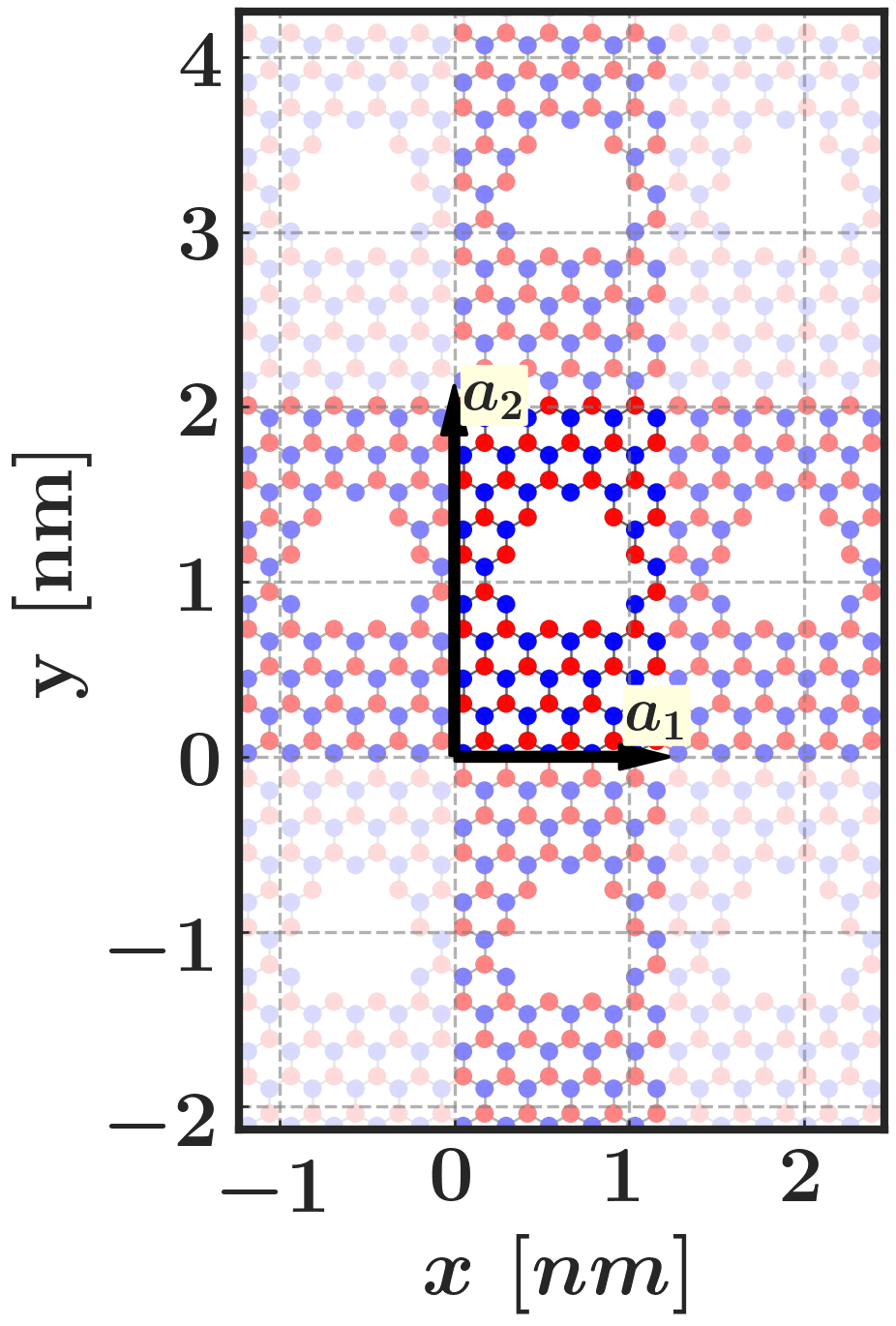}
	\caption{a) Unit cell of Holey Graphene (HG), where the lattice vectors are defined as $\boldsymbol{a}_{1}=n \boldsymbol{l}_{1}$ and $\boldsymbol{a}_{2}=m \boldsymbol{l}_{2}$, with $\boldsymbol{l}_{1}$ and $\boldsymbol{l}_{2}$ being the primitive vectors of a 4-atom graphene basis and $n,m \in \mathbb{Z}$. Additionally, atoms inside the circular region of radius $R$, highlighted in red, are removed. For notation purposes, we refer to this unit cell as $UCHG(n,m,R)$.  b) Periodic representation of HG, where sublattice $A$ sites are shown in blue and sublattice $B$ sites in red.}
	\label{fig:holey-graphene-model-lattice}
\end{figure}

The model is based on a graphene lattice with periodically distributed holes, known as Holey Graphene (HG) \cite{Champo_2024}. As illustrated in Fig. \ref{fig:holey-graphene-model-lattice} a),  the unit cell is constructed by defining the lattice vectors as $\boldsymbol{a}_{1}=n \boldsymbol{l}_{1}$ and $\boldsymbol{a}_{2}=m \boldsymbol{l}_{2}$, where $\boldsymbol{l}_{i}, i=1,2$ are the primitive vectors of a graphene basis consisting of four atoms. Within this unit cell, atomic sites inside a circular region of radius $R$ (\r{A}), centered at $(\boldsymbol{a}_{1}+\boldsymbol{a}_{2})/2$, are removed (marked in red). For notation purposes, we refer to this unit cell as $UCHG(n,m,R)$.

From Fig. \ref{fig:holey-graphene-model-lattice}b, it is clear that this construction maintains translational symmetry, enabling a simplified treatment using Bloch’s theorem.  As is well known, graphene is a bipartite lattice,i.e., is made from two trigonal sublattices, one displaced from the other \cite{Katsnelson_2012}. Usually, one of these sublattices is called $A$ and the other $B$. Following this convention and, unless stated otherwise, in the figures, sublattice $A$ sites will henceforth be shown in blue, while sublattice $B$ sites will be represented in red.

The tight-binding Hamiltonian describing this system is given by \cite{Champo_2024}:

\begin{equation} \label{eq:modelo_hamiltoniano_HG}
	\mathcal{H}= \sum_{\braket{ij}} t_{0} \hat{c}_{\boldsymbol{r}_{i}}^{\dagger} \hat{c}_{\boldsymbol{r}_{j}}+h.c.
\end{equation}

where $\braket{ij}$ denotes the summation over all nearest-neighbor sites of $\boldsymbol{r}_{i}$ with position $\boldsymbol{r}_{j}$ satisfying $|\boldsymbol{r}_{i}-\boldsymbol{r}_{j}|=a_0$, where $a_0=1.42$ \r{A} is the carbon-carbon bond length in graphene. The operators $\hat{c}_{\boldsymbol{r}_{i}}^{\dagger} (\hat{c}_{\boldsymbol{r}_{i}})$ correspond to the creation (annihilation) operators, and $t_0=-2.8$ eV represents the hopping integral between sites $i$ and $j$.

Additionally, the Hamiltonian operator in reciprocal space $\boldsymbol{k}$ can be constructed numerically. This operator depends on the number of sites in the unit cell and we denoted as $\hat{H}_{T}(\boldsymbol{k})$.


\section{Computational Methods} \label{sec:Numerical Methods}

In this section, we review the numerical methods employed to characterize certain optical properties. It is divided into two subsections: the first presents the optical conductivity, while the second covers the polarization function and the electron energy loss function.

\subsection{Optical Conductivities} \label{sec:Optical conductivities}

Physically, the optical conductivity relates an applied electric field to the resulting current in the material and can be calculated from the quantum mechanical transition rates between electronic states. The optical conductivity can be derived from the Kubo formula \cite{Kubo1957} and evaluated using the Kubo-Greenwood  expression \cite{CALDERIN2017118, Kuang2022, Tbplas2023},

\begin{eqnarray}\label{eq:definition of conductivity}
		\sigma_{\alpha_{1} \alpha_{2}}(\hbar \omega)= \frac{ie^{2}\hbar}{N_{k} \Omega_{c}} \sum_{\boldsymbol{k}} \sum_{n,m} \frac{f_{m \boldsymbol{k}}-f_{n \boldsymbol{k}}}{E_{m \boldsymbol{k}}-E_{n \boldsymbol{k}}} \nonumber \\
		 \times \frac{\braket{\psi_{n \boldsymbol{k}}|v_{\alpha_{1}}| \psi_{m\boldsymbol{k}}}\braket{\psi_{m \boldsymbol{k}}|v_{\alpha_{2}}| \psi_{n\boldsymbol{k}}}}{E_{m \boldsymbol{k}}-E_{n \boldsymbol{k}}-(\hbar \omega+i \eta^{+})} 
\end{eqnarray}

where $N_{\boldsymbol{k}}$ is the number of $\boldsymbol{k}$-points in the first Brillouin zone, $\Omega_{c}$ is the unit cell area, and $v_{\alpha_{i}}$ is the $\alpha_{i}$ component of the velocity operator. Additionally, $\psi_{n, \boldsymbol{k}}$ are the eigenstates of the Hamiltonian $\hat{H}_{T}(\boldsymbol{k})$, with eigenvalues $E_{n \boldsymbol{k}}$, and $f_{n \boldsymbol{k}}$ denotes the occupation number for the $n$-th band and momentum $\boldsymbol{k}$.

Combining the Kubo formula with the tight-binding propagation method (TBPM), the real part  of the optical conductivity (excluding the Drude contribution at $\omega=0$) is given by \cite{Tbplas2023,Kuang2022}

\begin{eqnarray}
	\label{eq:real-part-optical-conductivity-tbpm}
		\mathcal{R}[\sigma_{\alpha_{1} \alpha_{2}}](\hbar \omega)= \lim_{\varepsilon\to 0^{+}} \frac{e^{-\beta \hbar \omega}-1}{\hbar \omega \Omega} \int_{0}^{\infty} dt\, e^{-\varepsilon t} \sin(\omega t) \nonumber \\
		 \times 2 \mathcal{I}[\braket{\psi|f(\mathcal{H}) e^{i\mathcal{H}t}J_{\alpha_1}(1-f(\mathcal{H}))J_{\alpha_2}|\psi}]
\end{eqnarray}
where $\Omega$ is the system area, $\beta=1/k_{B}T$ with $k_B$ as the Boltzmann constant, $\mathcal{H}$ is the Hamiltonian \ref{eq:modelo_hamiltoniano_HG}, $f(\mathcal{H})$ is the Fermi distribution defined as

\begin{equation}
	\label{eq:fermi-dirac distribution}
	f(\mathcal{H})= \frac{1}{e^{\beta (\mathcal{H}-\mu)}+1},
\end{equation}

$J_{\alpha_i}$ denotes the $\alpha_i$-th component of the current density operator ${\boldsymbol{J}}$ \cite{Tbplas2023}

\begin{equation}
	\label{eq:current-density-operator}
	\boldsymbol{J}= -\frac{ie}{\hbar} \sum_{ij}  (\hat{\boldsymbol{r}}_{i}-\hat{\boldsymbol{r}}_{j})t_{ij}\hat{c}_{\boldsymbol{r}_{i}} \hat{c}_{\boldsymbol{r}_{j}},
\end{equation}

with $\hat{\boldsymbol{r}}$ as the position operator. Finally, the initial state $\ket{\psi}$ is a random superposition of all real-space basis states \cite{Kuang2022}

\begin{equation}
	\label{eq:random-state}
	\ket{\psi}= \sum_{i} a_{i} \ket{i},
\end{equation}

where $\{\ket{i}\}$ is the real-space basis set, and $a_{i}$ are normalized random complex coefficients ensuring $\sum_{i} |a_{i}|^2=1$, which guarantees the accuracy of the optical conductivity calculation \cite{Tbplas2023}.

In the other hand, the imaginary part of the optical conductivity is obtained via the Kramers-Kronig relation \cite{Tbplas2023},

\begin{equation}
	\label{eq:imaginary-part-optical-conductivity}
	\mathcal{I}[\sigma_{\alpha_1 \alpha_2}]=- \frac{1}{\pi} \mathcal{P} \int_{- \infty}^{\infty} \frac{\mathcal{R}[\sigma_{\alpha_1 \alpha_2}](\hbar \omega')}{\omega'-\omega} d \omega'.
\end{equation}

Additionally, the optical transmission for normally incident light is related to the optical conductivity \cite{Kuang2022}:

\begin{equation}
	\label{eq:optical-transmission}
	T=\left| 1+\frac{2\pi}{c} \sigma(\omega)\right|\approx 1-A,
\end{equation}
where $A$ is the normal-incidence absorbance, given by \cite{Kuang2022}:

\begin{equation}
	\label{eq:optical-absorbance}
	A = \frac{4\pi}{c} \mathcal{R}[\sigma(\omega)].
\end{equation}

As can be observed from Eq. \ref{eq:real-part-optical-conductivity-tbpm}, the TBPM method allows the computation of optical conductivities with linear scaling $\mathcal{O}(N)$ in the number of real-space states $N$. In contrast, exact diagonalization in Eq. \ref{eq:definition of conductivity} scales as $\mathcal{O}(N^2)$. Therefore, the TBPM facilitates the calculation of optical conductivity for large-scale periodic structures \cite{Kuang2022, Tbplas2023}.

\subsection{Polarization, Dielectric, and Electron Energy Loss Functions} \label{sec:plasmons}

The polarization function, $\boldsymbol{\Pi}_{0}$, also known as the charge susceptibility or the non-interacting density-density response function, describes charge fluctuations and single-particle transitions. It plays a crucial role in the description of collective excitations and screening in materials. The polarization function can be evaluated using the Lindhard function, given by \cite{Kuang2022, Tbplas2023, mahan2013many, Coleman_2015, Giuliani_Vignale_2005},

\begin{eqnarray} \label{eq:lindhard_function_definition}
		\boldsymbol{\Pi}_{0}(\boldsymbol{q}, \omega)= -\frac{g_{s}}{(2 \pi)^{D}} \int_{BZ} d^{D} \boldsymbol{k} \sum_{m,n}  \left[ \frac{f_{m \boldsymbol{k}}-f_{n \boldsymbol{k+q}}}{E_{m\boldsymbol{k}}-E_{n\boldsymbol{k+q}}+\hbar \omega +i \eta}\right. \nonumber \\
	\left.	\times |\braket{\psi_{n \boldsymbol{k+q}}|e^{i \boldsymbol{q\cdot r}}|\psi_{m\boldsymbol{k}}}|^{2}\right]
\end{eqnarray}
where $g_s$ is the spin degeneracy, $D$ is the system dimensionality, and $\eta \to 0^{+}$. By combining it with the TBPM, such a polarization function can be rewritten as \cite{Kuang2022, Tbplas2023}

\begin{eqnarray}
	\label{eq:polarization_function_tbpm}
		\boldsymbol{\Pi}_{0}(\boldsymbol{q},\omega)= -\frac{2}{\Omega} \int_{0}^{\infty} dt \, e^{i \omega t} \mathcal{I}\left[ \langle \psi | f(\mathcal{H} e^{i \mathcal{H} t}) \rho(\boldsymbol{q}) \right. \nonumber \\
		\left. \times e^{-i \mathcal{H} t}  (1-f(\mathcal{H})) \rho(- \boldsymbol{q}) | \psi\rangle\right]
\end{eqnarray}
where

\begin{equation}
	\label{eq:density_operator}
	\rho(\boldsymbol{q})= \sum_{i} \hat{c}_{\boldsymbol{r}_{i}}^{\dagger} \hat{c}_{\boldsymbol{r}_{j}} \exp(i \boldsymbol{q \cdot r}_{i})
\end{equation}
is the density operator. The dynamic dielectric function $\epsilon(\boldsymbol{q}, \omega)$ can be derived from $\boldsymbol{\Pi}_{0}$ within the random phase approximation (RPA), and is given by \cite{jones1973theoretical}

\begin{equation} \label{eq:dielectric_function_definition}
	\epsilon(\boldsymbol{q},\omega)= 1- V(\boldsymbol{q}) \boldsymbol{\Pi}_{0}(\boldsymbol{q}, \omega),
\end{equation}
where $V(\boldsymbol{q})$ is the Fourier component of the Coulomb potential. For 2D systems \cite{Kuang2022, Tbplas2023},

\begin{equation} \label{eq:Coulomb_potential}
	V(\boldsymbol{q})= \frac{2 \pi e^{2}}{\kappa |\boldsymbol{q}|},
\end{equation}
where $\kappa$ is the background dielectric constant. In the long-wavelength limit ($\boldsymbol{q} \to 0$), the RPA dielectric function is related to the optical conductivity $\sigma(\omega)$ by \cite{jones1973theoretical,Kuang2022,Tbplas2023, Nazarov_2015}.

\begin{equation}
	\label{eq:dielectric_function_and_optical_conductivity}
	\epsilon(\boldsymbol{q}, \omega)= 1+ \frac{i |\boldsymbol{q}|^2 V(\boldsymbol{q})}{\omega} \sigma(\omega).
\end{equation}

Finally, the electron energy loss function is given by \cite{Kuang2022, Tbplas2023}

\begin{equation} \label{eq:definition_of_loss_function}
	\mathcal{L}(\boldsymbol{q}, \omega)= - \mathrm{Im}\left[ \frac{1}{\epsilon(\boldsymbol{q}, \omega)} \right].
\end{equation}

Experimentally, the electron energy loss function can be measured using electron energy loss spectroscopy (EELS).  A well-defined plasmonic mode, characterized by frequency $\omega_P$ and wave vector $\boldsymbol{q}$, exists when $\mathcal{L}(\boldsymbol{q}, \omega_P)$ exhibits a peak, indicating that the function has a pole \cite{Tbplas2023}. Equivalently, this condition is satisfied when $\epsilon(\boldsymbol{q},\omega_P)=0$ \cite{jones1973theoretical}. It is important to note that the imaginary part of the polarization function, consistently denoted in this work as $\mathcal{I}[\boldsymbol{\Pi}_{0}]$, is fundamental for identifying the regions of the particle-hole continuum. Physically, a non-zero value of $\mathcal{I}[\boldsymbol{\Pi}_{0}(\boldsymbol{q}, \omega)]$ indicates the possibility for a plasmon with momentum $\boldsymbol{q}$ and energy $\hbar\omega$ to decay into an electron-hole pair excitation, a process known as Landau damping. Therefore, the peaks in the energy loss function $\mathcal{L}(\boldsymbol{q}, \omega)$ that lie outside these regions (where $\mathcal{I}[\boldsymbol{\Pi}_{0}] \approx 0$) correspond to well-defined, long-lived collective plasmons \cite{Pines1962, Giuliani_Vignale_2005}.


\section{Results and Discussion}
\label{sec:Results and Discussion}

In this section, we present and analyze the results obtained from our study of holey graphene (HG). We focus on understanding how the introduction of holes affects the material’s optoelectronic properties by examining its optical conductivity and plasmonic dispersion.

\begin{figure*}[t]
	\centering
	\large a)\includegraphics[height=0.18\textheight]{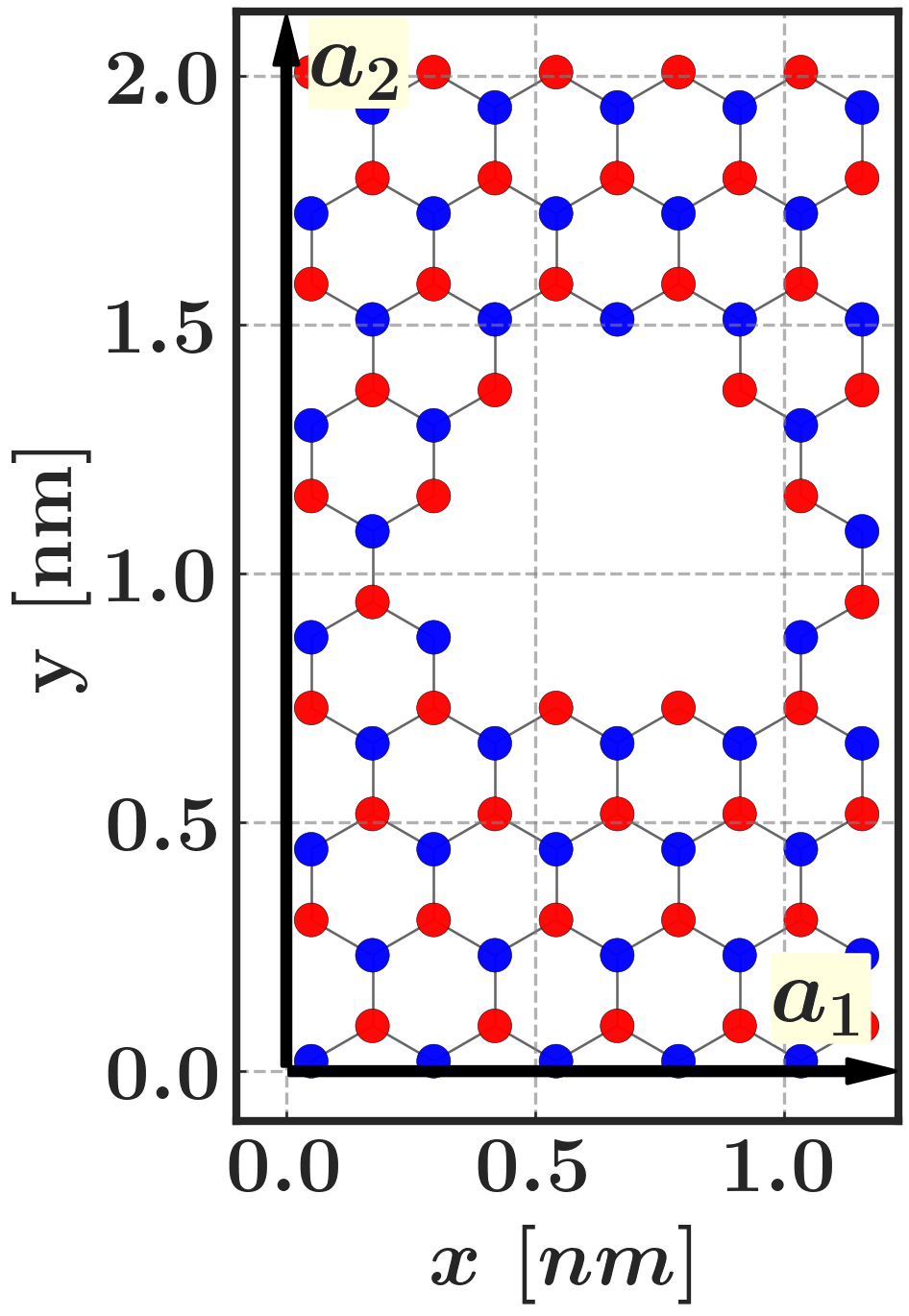}
	\large b)\includegraphics[height=0.18\textheight]{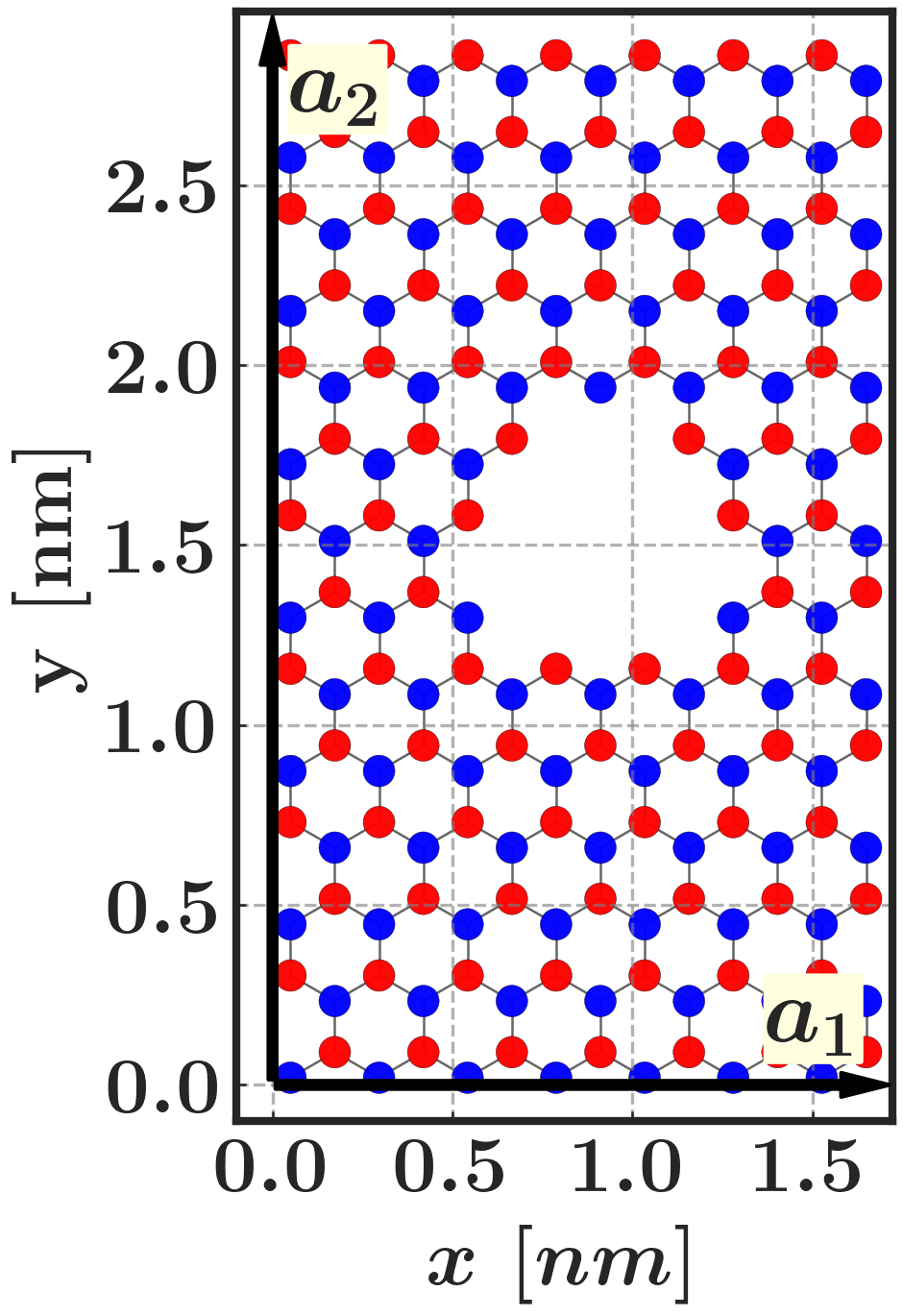}\\
	\large c)\includegraphics[height=0.18\textheight]{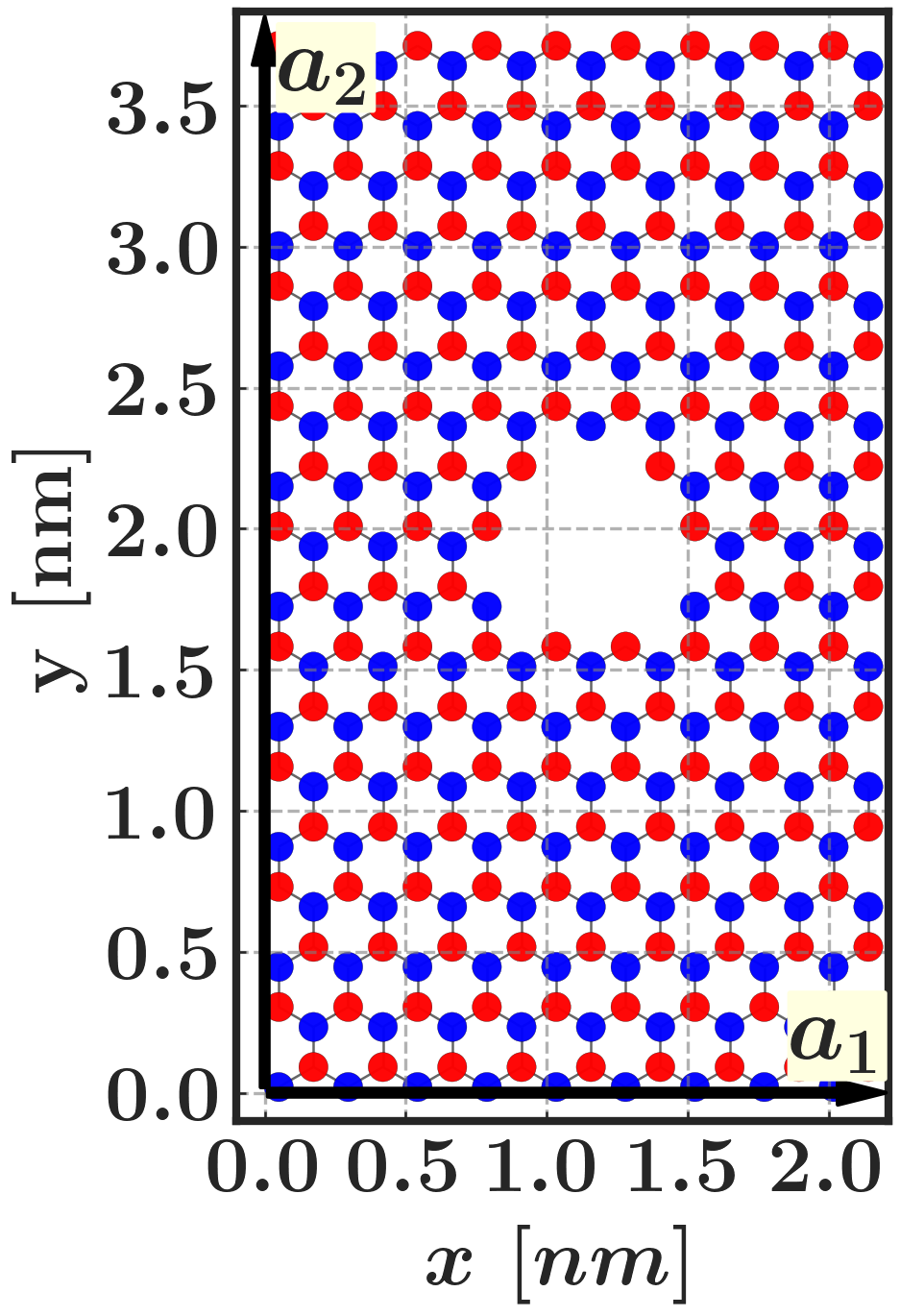}
	\large d)\includegraphics[height=0.18\textheight]{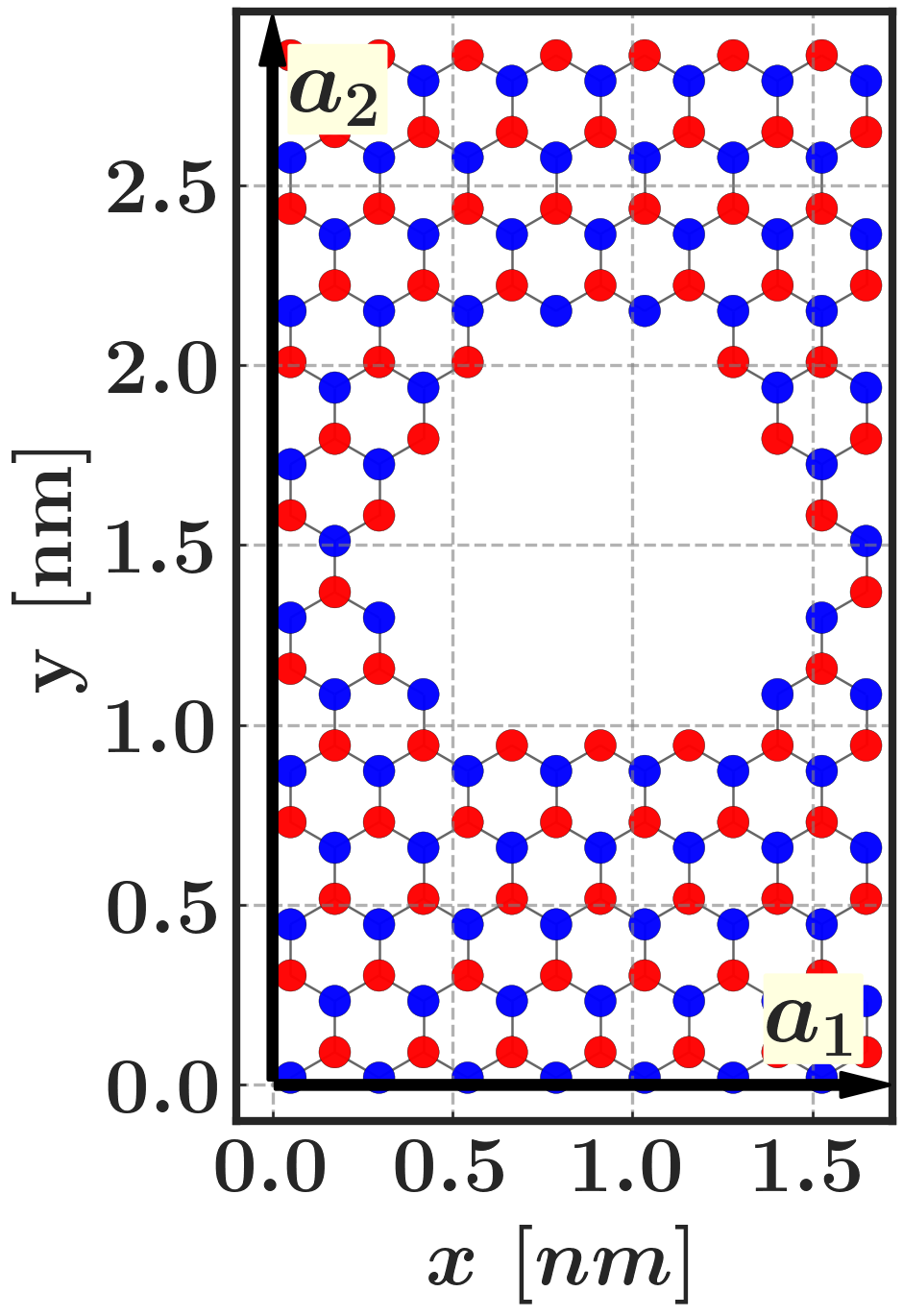}
	\large e)\includegraphics[height=0.18\textheight]{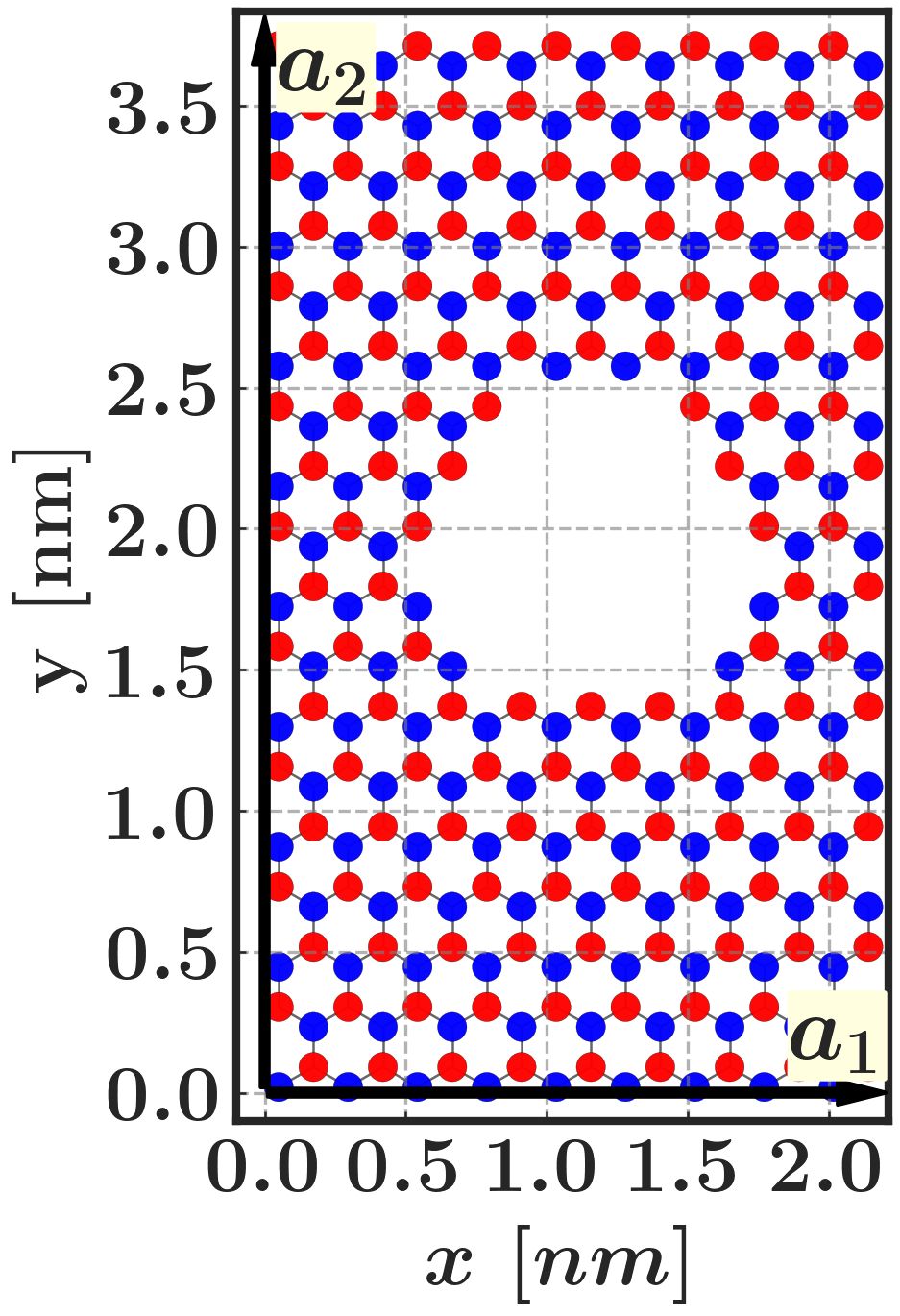}
	\caption{Illustration of five unit cells with varying hole radii and lattice constants for (a) UCHG$(5,5,3)$, (b) UCHG$(7,7,3)$, (c) UCHG$(9,9,3)$, (d) UCHG$(7,7,5.2)$, and (e) UCHG$(9,9,5.2)$. Atom colors denote the two bipartite sublattices, $A$ and $B$.}
	\label{fig:2-1}
\end{figure*}

Figure~\ref{fig:2-1} illustrates the five unit cells of the systems under investigation: UCHG$(5,5,3)$, UCHG$(7,7,3)$, UCHG$(9,9,3)$, UCHG$(7,7,5.2)$, and UCHG$(9,9,5.2)$. We place special emphasis on UCHG$(5,5,3)$; the corresponding plots for the other systems are provided in Appendix~\ref{app:Holey graphene}. {In this section, we first analyze the plasmon dispersion, revealing the emergence of quasi-flat and hyperbolic modes. We then connect these features to the calculated optical conductivity, demonstrating a strong anisotropy directly linked to the hole-induced symmetry breaking.} In the following subsections we discuss their corresponding optical properties.

\subsection{Plasmon Dispersion and Hyperbolic Plasmons}

Plasmons, which are collective oscillations of free charge carriers, are crucial for understanding the material's response to electromagnetic fields \cite{Zhu2014,Grankin2023}.  As we shall discuss in this subsection, the introduction of periodic holes into the graphene lattice leads to significant modifications in the dispersion of plasmons compared to pristine graphene. The reason is the appearance of flat bands in the electronic structure \cite{Champo_2024}, a  characteristic of many 2D materials with broken path-exchange symmetry \cite{Jun2023,Espinosa-Champo_2024,Champo_2024}. We shall first analyze the plasmon dispersion for the $UCHG(5,5,3)$ system.

\begin{figure}[t]
	\centering
	\includegraphics[scale=0.46]{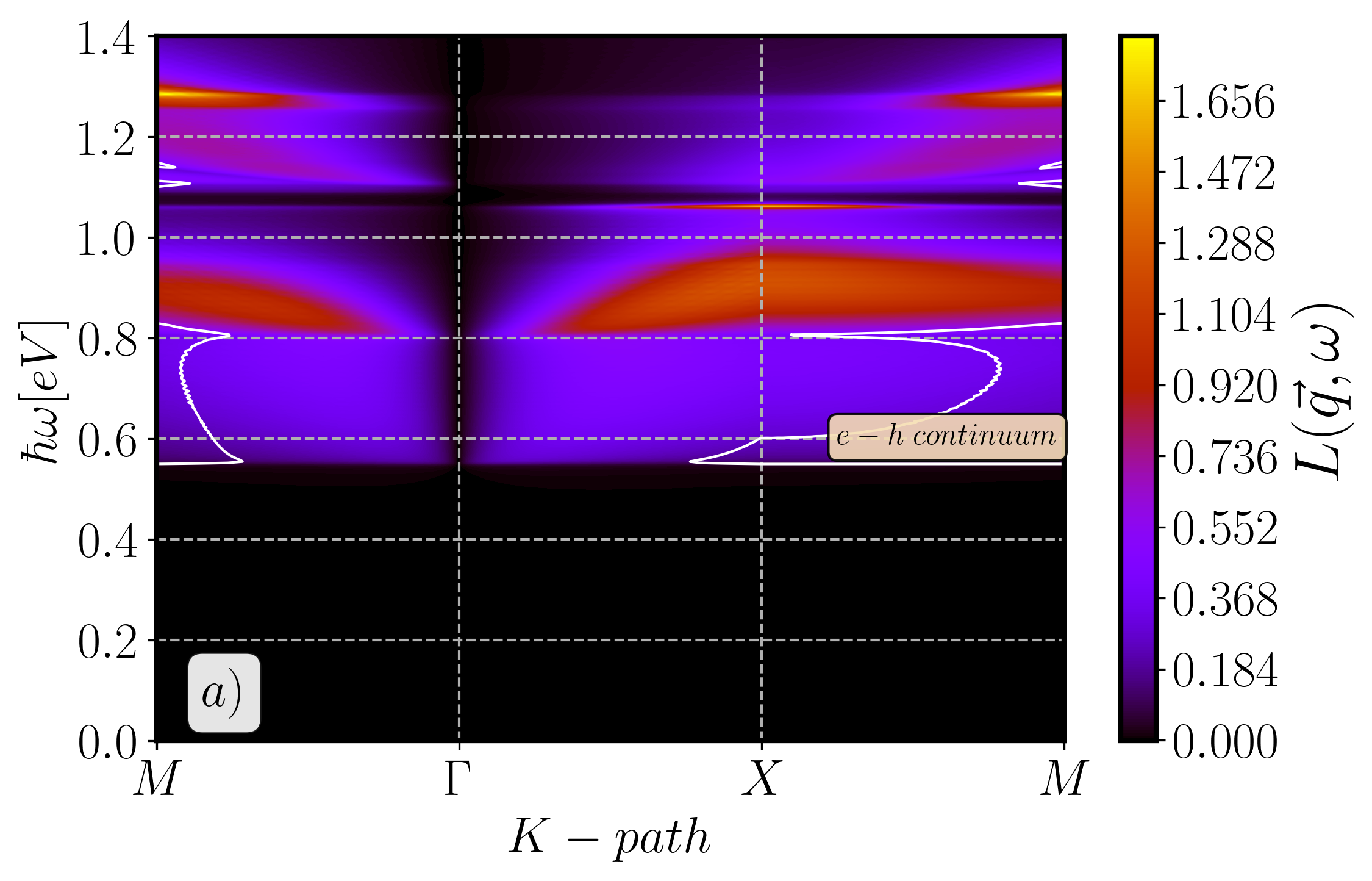}\\
	\includegraphics[scale=0.46]{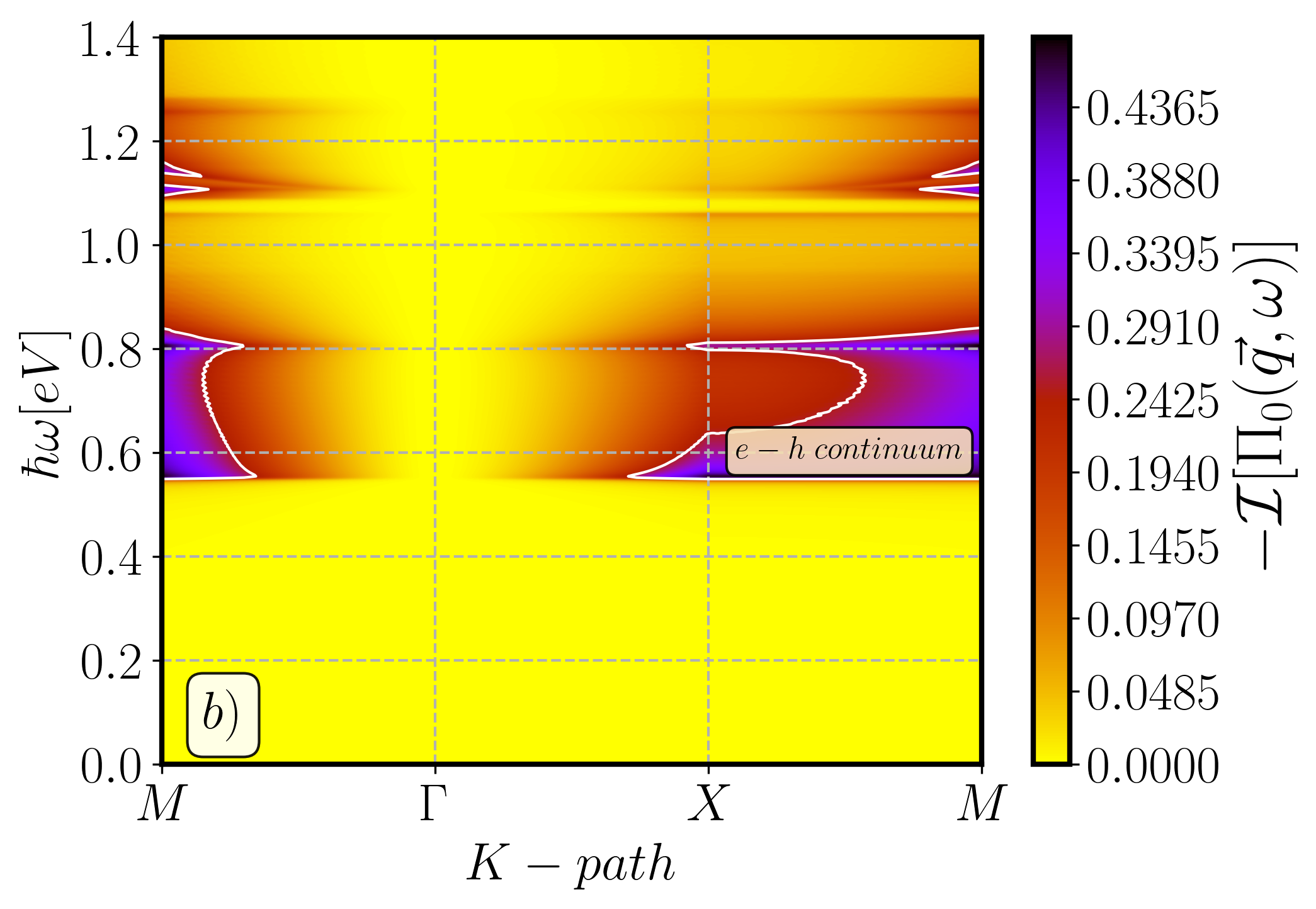}
	\caption{(a) Plasmon dispersion branches for UCHG$(5,5,3)$, showing quasi-flat modes arising from the flat electronic band. (b) Negative imaginary part of the polarization function, $-\mathcal{I}[\Pi_0(\mathbf{q},\omega)]$, delineating the particle-hole continuum (white lines) where Landau damping occurs. Most plasmon modes lie outside this continuum, indicating long lifetimes.}
	\label{fig:plasmons-branches-loss-function-553}
\end{figure}

In Fig. \ref{fig:plasmons-branches-loss-function-553} a), we observe the formation of quasi-flat plasmonic bands.  These nearly flat plasmonic bands are especially noticeable in configurations with larger hole radii and higher periodicity, see App. \ref{app:Holey graphene}.

On the other hand, the polarization function is particularly important for understanding the interaction between plasmons and the electron-hole continuum. In holey graphene, we find that the polarization function exhibits significant peaks at certain frequencies, corresponding to plasmonic excitations. These peaks are damped by the Landau damping process, where the plasmon modes decay into electron-hole pairs {\color{red}CITAS} \cite{Pines1962, Giuliani_Vignale_2005}. The regions where the Landau damping is significant are clearly visible in the imaginary part of the polarization function, which shows a negative value at certain frequencies. These damping regions are most prominent in systems with larger holes, suggesting that the hole configuration plays a key role in determining the lifetime of plasmonic modes.

{Thus, Fig. \ref{fig:plasmons-branches-loss-function-553}(b) shows the negative of the imaginary part of the polarization function, $-\mathcal{I}[\Pi_0(\mathbf{q},\omega)]$. As explained in the previous section, this quantity delineates the continuum of electron-hole pair excitations, which is the region where plasmons undergo Landau damping.} Comparing Figs. \ref{fig:plasmons-branches-loss-function-553} a) and b), we see highly tunable long-lived plasmons which are attributed to the coupling of the plasmonic modes with the flat bands.

\subsection{Optical Conductivity and Anisotropy}

\begin{figure*}[t]
	\centering
	\includegraphics[scale=0.5]{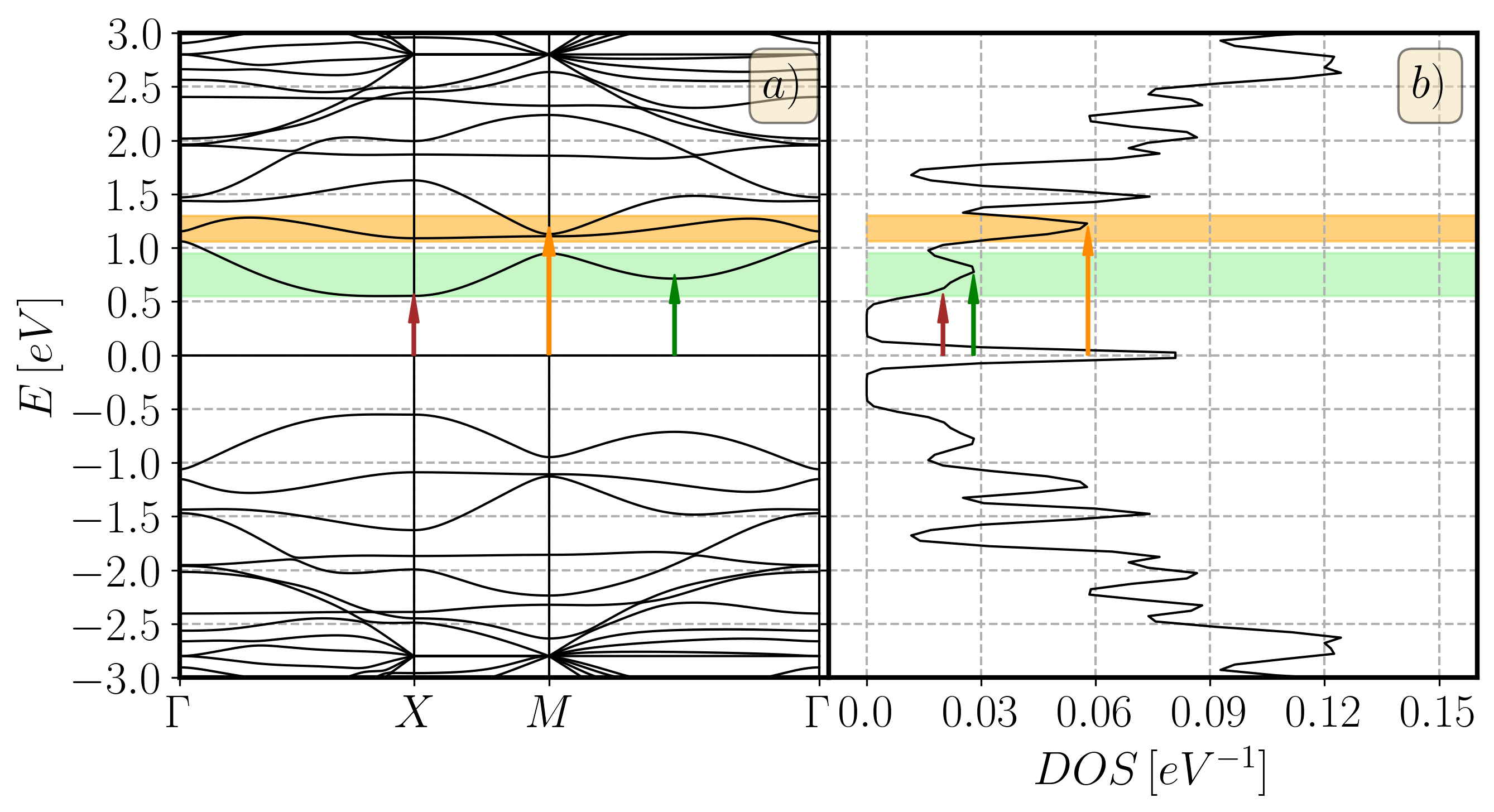}\\
	\includegraphics[scale=0.5]{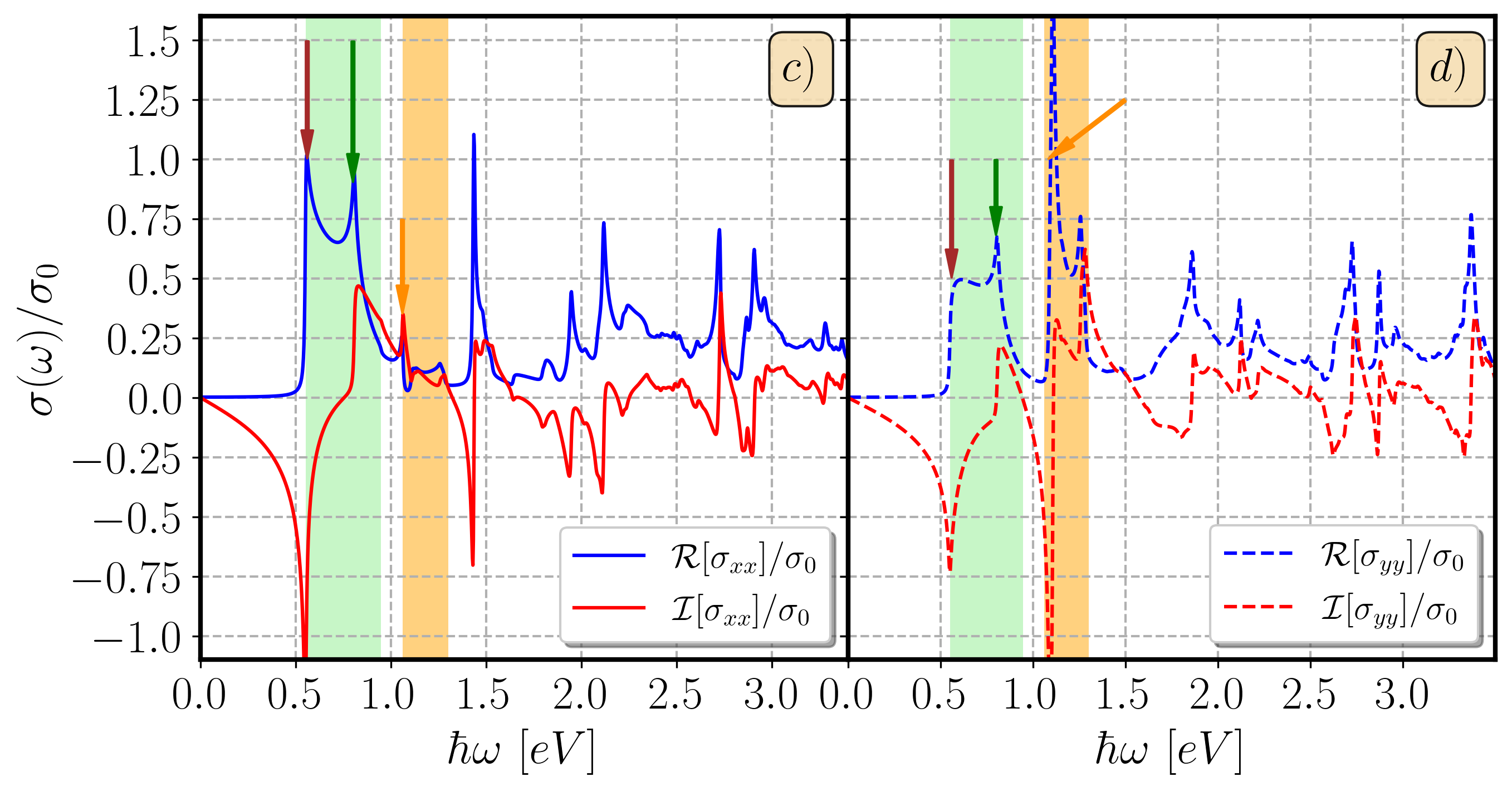}
	\caption{(a) Electronic band structure and (b) density of states (DOS) for UCHG$(5,5,3)$. (c) Real and (d) imaginary parts of the longitudinal optical conductivities, $\sigma_{xx}$ (solid lines) and $\sigma_{yy}$ (dashed lines). The blue and red line color denote the real and imaginary parts, respectively. Two energy regions are highlighted: the first (light green) corresponds to transitions near the first DOS peak, while the second (orange) corresponds to the second DOS peak. The anisotropic response arises because the higher-energy transitions starting at the flat band modes, occurring at the $X$ and $M$ points with $E\approx1.1\,$eV, predominantly affects $\sigma_{yy}$.  {The colored arrows in the band structure diagram would illustrate the primary electronic transitions. A green arrow would denote low-energy transitions near the $\Gamma$ point, which are responsible for the first absorption peak (light green region). An orange arrow would indicate transitions from the nearly flat band at the $X$ and $M$ points to higher conduction bands, leading to the second main absorption peak (orange region) and the pronounced anisotropy. A red arrow would point to a distinct intermediate-energy transition.}}
	\label{fig:bands-dos-holey-553}
\end{figure*}

{Having established the unique plasmonic behavior of HG, we now turn to the optical conductivity, which provides further insight into the anisotropic nature of the electron dynamics underlying these collective modes.} Figure~\ref{fig:bands-dos-holey-553} shows the longitudinal optical conductivity components, $\sigma_{xx}$ and $\sigma_{yy}$ of the UCHG$(5,5,3)$ system, computed via the Kubo–Greenwood formula combined with the tight-binding propagation method (TBPM; see Eqs.~\ref{eq:real-part-optical-conductivity-tbpm} and \ref{eq:imaginary-part-optical-conductivity}). We omit the transverse components, $\sigma_{xy}$ and $\sigma_{yx}$, as they are negligible.

The first notable feature in Figs.~\ref{fig:bands-dos-holey-553}(c) and (d) is the presence of three peaks (indicated by colored arrows), each arising from one of the principal electronic transitions above $E=0$. These transitions correspond to the first three bands visible in Fig.~\ref{fig:bands-dos-holey-553}(a), leading to the sharp peaks seen in the density of states (DOS) presented in Fig.~\ref{fig:bands-dos-holey-553} (b). This behavior is characteristic of gapped holey graphene systems: the peak positions shift depending on the band separations, as seen in UCHG$(7,7,3)$ and UCHG$(7,7,5.2)$ (Figs.~\ref{fig:bands-dos-holey-773} and \ref{fig:bands-dos-holey-775}). In contrast, UCHG$(9,9,3)$ and UCHG$(9,9,5.2)$ preserve a Dirac-like cone at the $\Gamma$ point due to local symmetries \cite{Abdiel2024}, resulting in a continuum of transitions that produce a step-like feature in the optical conductivity at low energies, similar to pristine graphene (Figs.~\ref{fig:bands-dos-holey-993} and \ref{fig:bands-dos-holey-995}).

\begin{figure*}[t]
	\centering
	\includegraphics[scale=0.5]{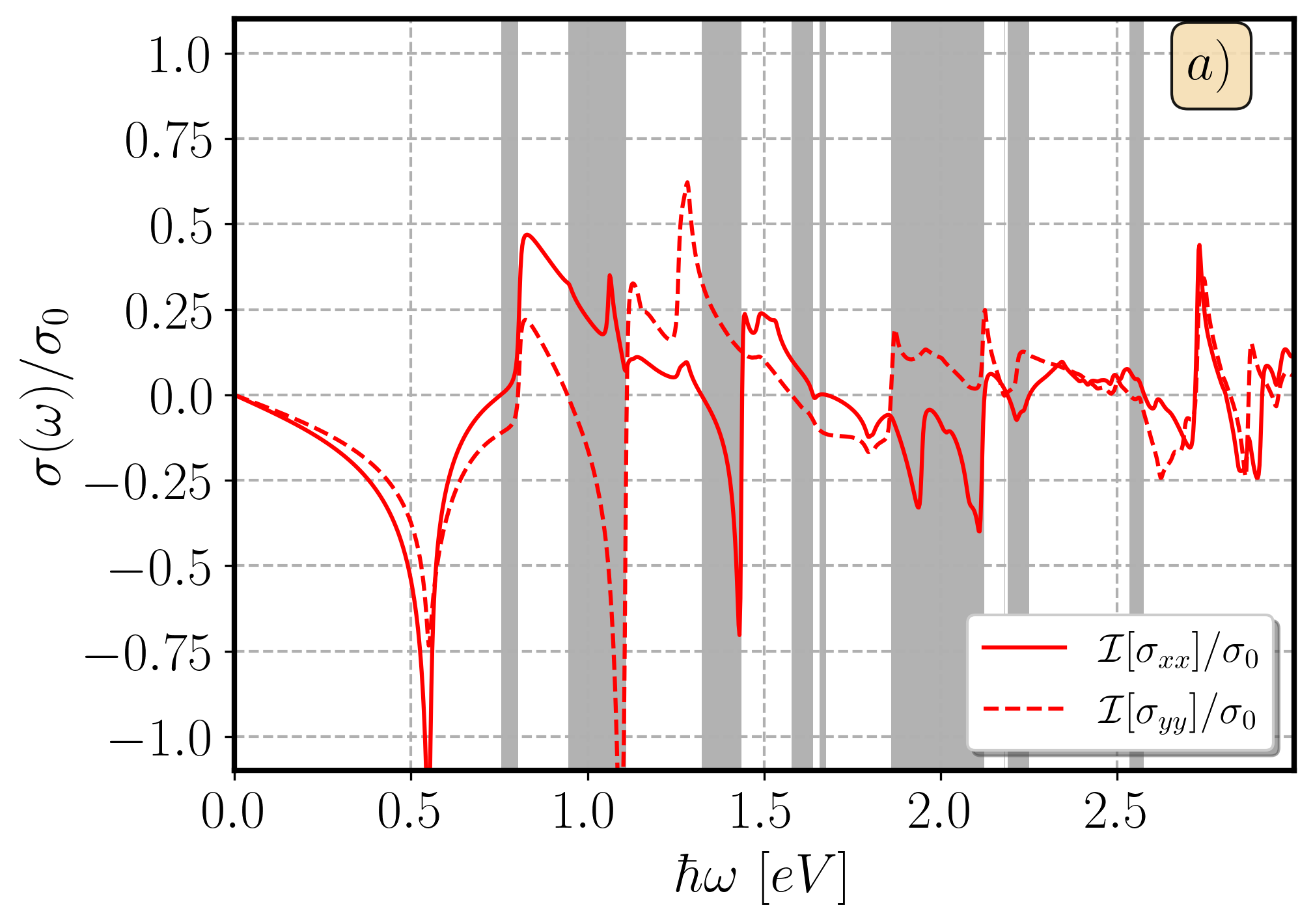}
	\includegraphics[scale=0.5]{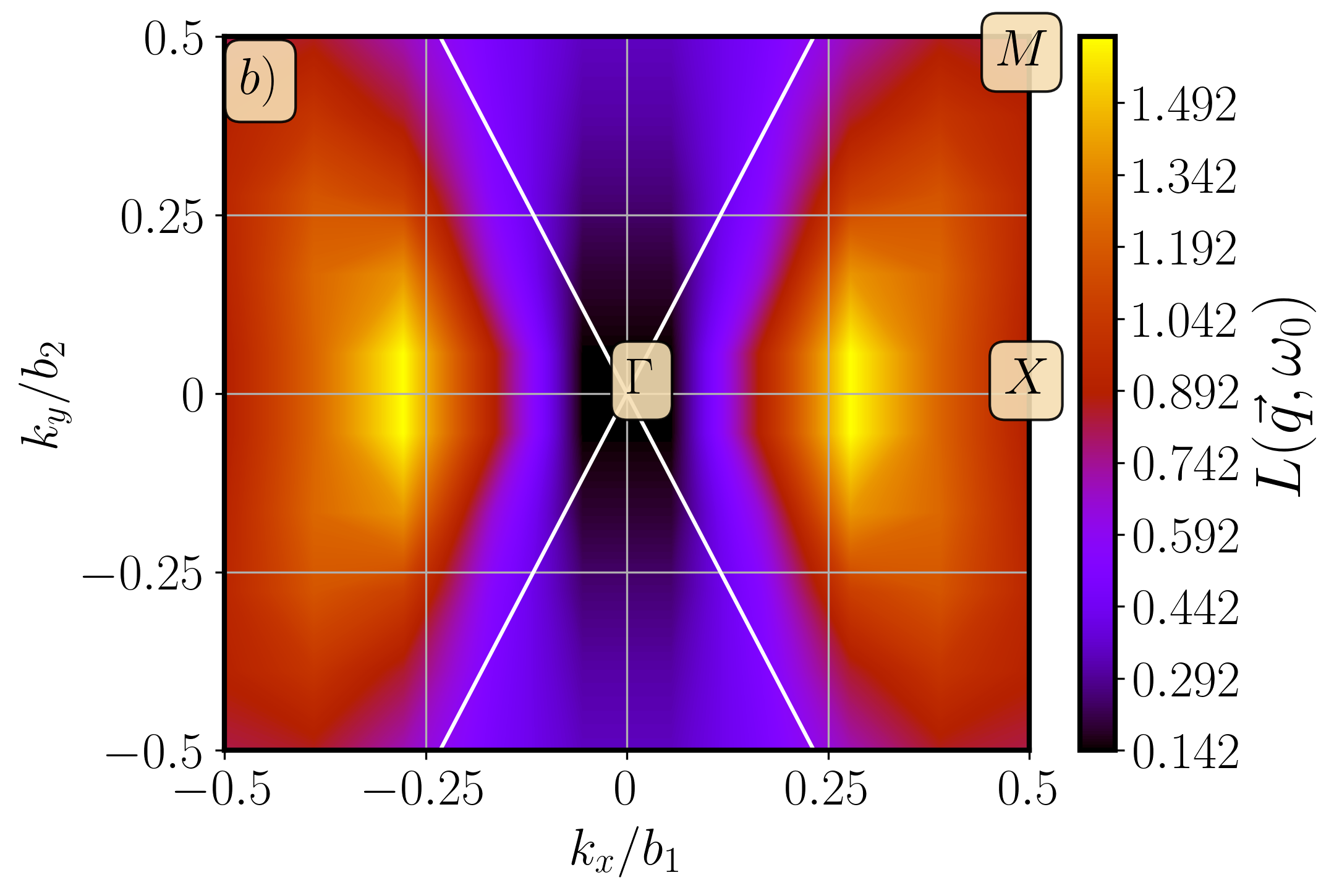}
	\caption{(a) Imaginary parts of $\sigma_{xx}$ (solid red) and $\sigma_{yy}$ (dashed red). Gray shading marks frequencies where $\mathcal{I}[\sigma_{xx}(\omega)]\,\mathcal{I}[\sigma_{yy}(\omega)]<0$, indicating hyperbolic plasmon support; white regions indicate elliptical plasmons. (b) Density plot of the electron energy loss function $S(\mathbf{q},\omega_0)$ at fixed $\hbar\omega_0=0.945\,$eV (second gray region), showing hyperbolic dispersion with asymptotes $k_y=\pm k_x[|\mathcal{I}[\sigma_{xx}]/\mathcal{I}[\sigma_{yy}]|]^{1/2}$. Here, $b_1$ and $b_2$ are the magnitudes of the reciprocal lattice vectors.}
	\label{fig:hyperbolic-loss-function-553}
\end{figure*}

As seen in Eq. \ref{eq:optical-absorbance}, the real part of the optical conductivity corresponds to the absorption of electromagnetic radiation, while the imaginary part relates to the energy dissipation in the material. In holey graphene, we observe sharp peaks in both components at specific energies, corresponding to the transitions between the first, second, and third electronic bands. These transitions are highly sensitive to the hole configuration, with larger holes leading to more pronounced peaks in the absorption spectrum.

Furthermore, the optical transmission for normally incident light is related to the optical conductivity via the Eq. \ref{eq:optical-transmission}. We find that holey graphene shows strong optical absorption in certain energy ranges, particularly when the plasmon modes are resonant with the incident light. This could enable the development of efficient light-harvesting devices based on holey graphene.

On the other hand, breaking the sublattice symmetry by introducing holes leads to a pronounced anisotropy in the optical response. As an example of such an effect, consider the UCHG$(5,5,3)$ system. Fig.~\ref{fig:bands-dos-holey-553}(c)–(d) reveals that its corresponding $\sigma_{xx}$ and $\sigma_{yy}$ differ substantially, which in turn distorts the electronic structure and transition probabilities. As seen in Fig. \ref{fig:bands-dos-holey-553}, this anisotropy response arises because the higher energy transition that occurs from the flat band at the $X$ and $M$ points, occurring at $E\approx1.1\,$eV, predominantly affects $\sigma_{yy}$. 

Figure~\ref{fig:hyperbolic-loss-function-553}(a) further highlights this anisotropy by comparing $\mathcal{I}[\sigma_{xx}(\omega)]$ and $\mathcal{I}[\sigma_{yy}(\omega)]$. Shaded gray areas satisfy
\begin{equation}
	\mathcal{I}[\sigma_{xx}(\omega)]\,\mathcal{I}[\sigma_{yy}(\omega)]<0,
	\label{eq:cond-hyperbolic}
\end{equation}
indicating frequency ranges that support hyperbolic plasmon propagation \cite{EdovanVeen2019, Nemilentsau2016}. In contrast, frequencies outside these regions support elliptical plasmons.

The emergence of hyperbolic plasmon modes is a key feature of holey graphene. In systems like UCHG(5,5,3), we observe that the plasmon dispersion becomes hyperbolic, with the plasmon frequency diverging as the wavevector increases. This behavior is similar to what is observed in hyperbolic metamaterials, where plasmon modes can be confined to extremely small spatial regions {\color{red} CITAS}\cite{Poddubny2013,Low2014}. The hyperbolic nature of the plasmon dispersion suggests that holey graphene could be used to design devices that manipulate light at subwavelength scales. This is particularly promising for applications in nanophotonics, where controlling plasmonic modes at the nanoscale is critical.

\section{Conclusions 
	\label{sec:conclusions}} 


Our findings demonstrate that holey graphene is a highly tunable platform for engineering anisotropic and hyperbolic plasmonic responses, opening new avenues for nanophotonic applications. We have shown that the plasmonic modes are highly sensitive to the hole geometry. Specifically, increasing the hole radius tends to lower the plasmon frequency, while tighter periodic arrangements lead to higher-frequency modes. Crucially, we found that the same hole-induced sublattice symmetry breaking that generates electronic flat bands is also responsible for the emergence of nearly flat plasmonic bands, hyperbolic plasmons, and a marked anisotropy in the optical conductivity.

These tuning effects are crucial for tailoring the plasmonic response of holey graphene for specific applications, such as sensors or modulators. By varying the hole size and periodicity, one can precisely control the plasmonic properties of the material, making it a versatile platform for plasmonic applications.

When comparing holey graphene to other 2D materials, such as twisted bilayer graphene (tBLG) and graphene antidot lattices, we observe several similarities and differences. Both tBLG and holey graphene exhibit the formation of flat bands, which are associated with the emergence of exotic plasmonic and electronic properties. However, holey graphene offers a unique advantage in terms of simplicity and tunability. Unlike tBLG, which requires specific twists to induce flat bands, holey graphene can be engineered by simply adjusting the hole pattern, making it easier to control the material's electronic and optical properties.

While the results presented here are based on numerical simulations, experimental validation is crucial for confirming the predicted plasmonic and optical behaviors. Future experimental studies should focus on fabricating holey graphene using techniques such as nanopore lithography or chemical vapor deposition. By measuring the optical conductivity, plasmon dispersion, and polarization functions experimentally, we can compare the theoretical predictions with real-world data, providing a more comprehensive understanding of the material's properties.

In the future, we will also explore the effects of different hole patterns on other material properties, such as thermal conductivity, electron mobility, and magnetoresistance. Additionally, combining holey graphene with other 2D materials, such as transition metal dichalcogenides or topological insulators, could open new avenues for designing novel hybrid materials with enhanced optical and electronic properties. The integration of holey graphene with quantum technologies could also lead to new devices for quantum information processing and highly sensitive sensors based on plasmon resonance shifts.

\ack
AJEC and GGN thanks the CONAHCyT fellowship (No. CVU 1007044) and the Universidad Nacional Autónoma de México (UNAM) for providing financial support (UNAM DGAPA PAPIIT IN101924 and CONAHCyT project 1564464). The authors acknowledge and express gratitude to Ram\'on Carrillo-Bastos for enlightening discussions on the hyperbolic plasmons phenomena in 2D materials. We also thanks to Carlos Ernesto L\'opez Natar\'en from Secretaria T\'ecnica de C\'omputo y Telecomunicaciones for his valuable support to implement high-performance numerical calculations.

\appendix

\section{Other holey graphene systems \label{app:Holey graphene}}

\subsection{Holey graphene 773}

\begin{figure*}[h]
	\centering
	\includegraphics[scale=0.5]{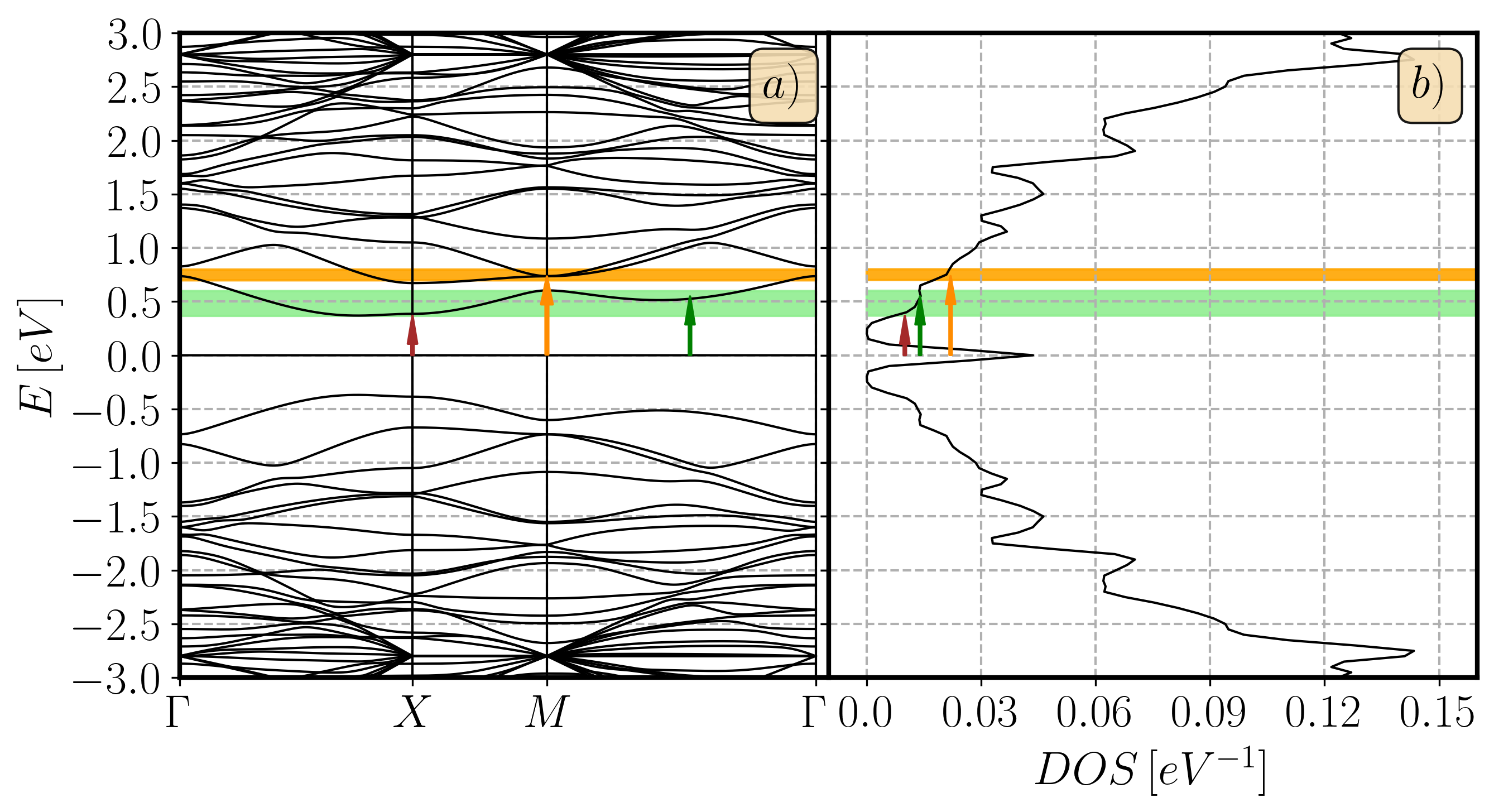}\\
	\includegraphics[scale=0.5]{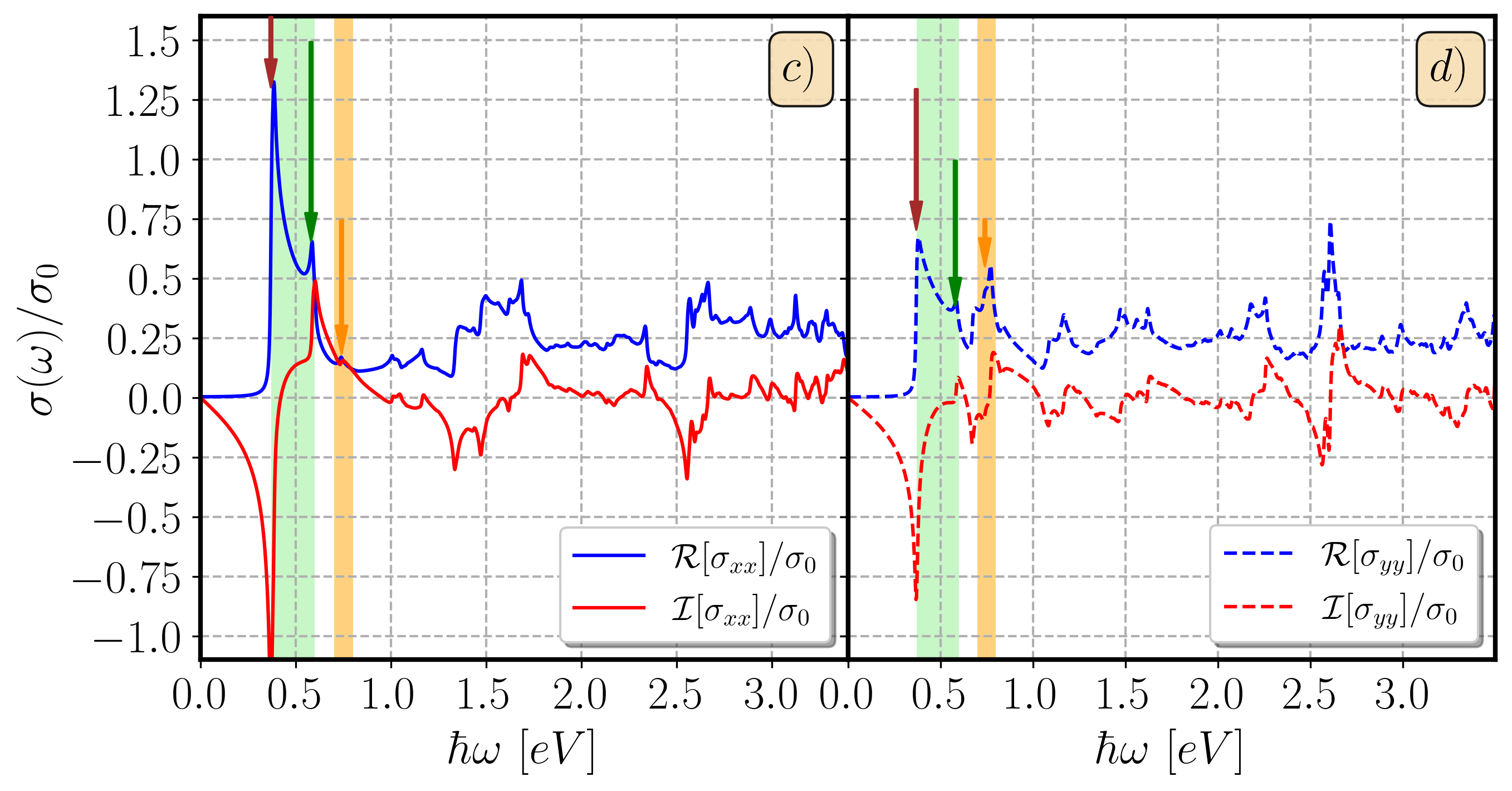}
	\caption{{(a) Electronic band structure and (b) density of states (DOS) for the gapped $UCHG(7,7,3)$ system. (c) and (d) show the corresponding longitudinal optical conductivities. The key feature is the strong anisotropy between $\sigma_{xx}$ (solid lines) and $\sigma_{yy}$ (dashed lines), especially in the orange-highlighted region around $1.1\,$eV. This anisotropy originates from electronic transitions involving the flat band sections near the $X$ and $M$ points, which selectively enhance the optical response along the $y$-direction.}}
	\label{fig:bands-dos-holey-773}
\end{figure*}

\begin{figure*}[h]
	\centering
	\includegraphics[scale=0.5]{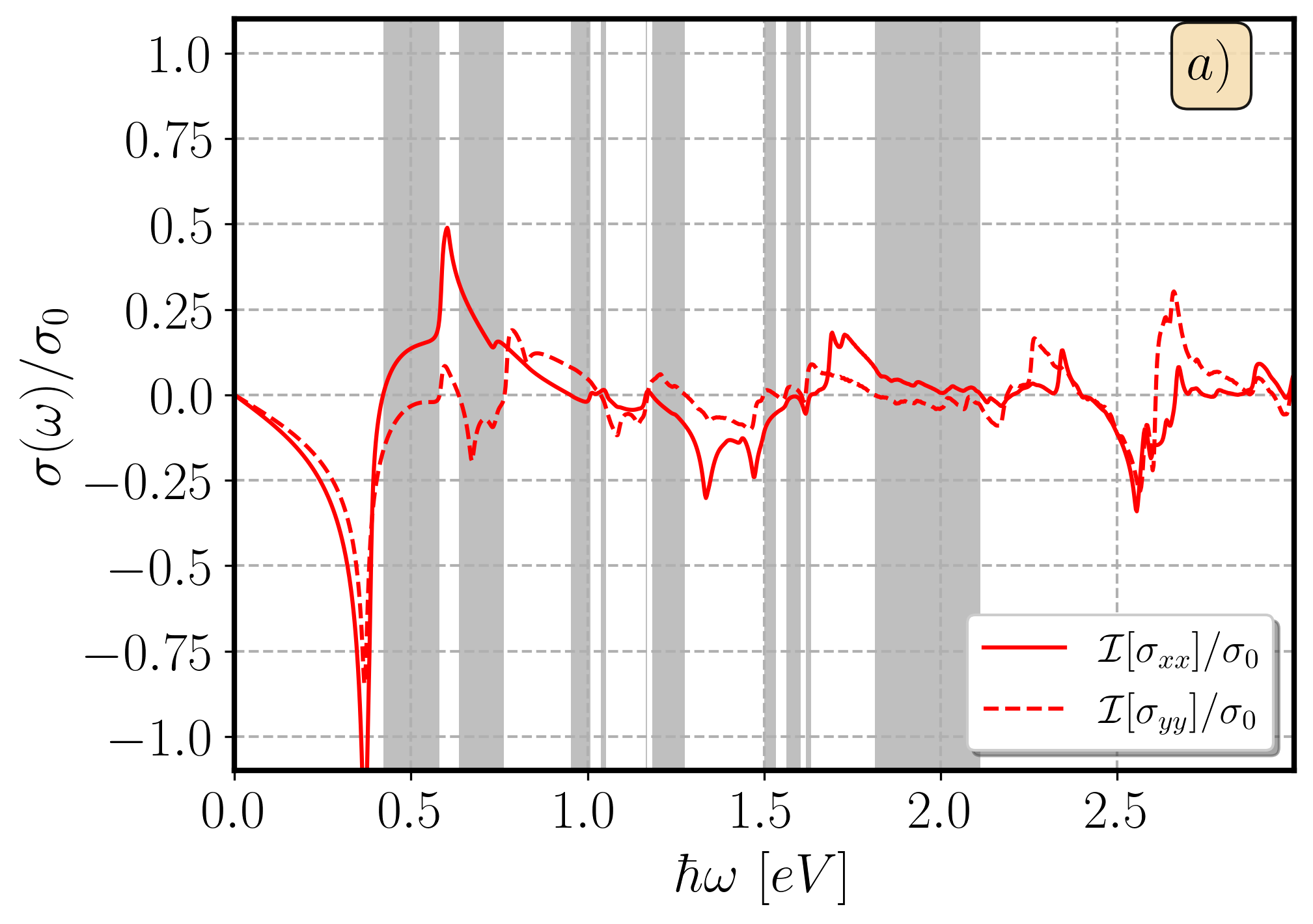}
	\caption{{Frequency dependence of the imaginary part of the optical conductivity for $UCHG(7,7,3)$. The shaded gray areas highlight the spectral windows where the condition for hyperbolic plasmons, $\mathcal{I}[\sigma_{xx}] \mathcal{I}[\sigma_{yy}] < 0$, is met. The material supports elliptical plasmons in the unshaded regions.}}
	\label{fig:hyperbolic-loss-function-773}
\end{figure*}

\begin{figure}[h]
	\centering
	\includegraphics[scale=0.5]{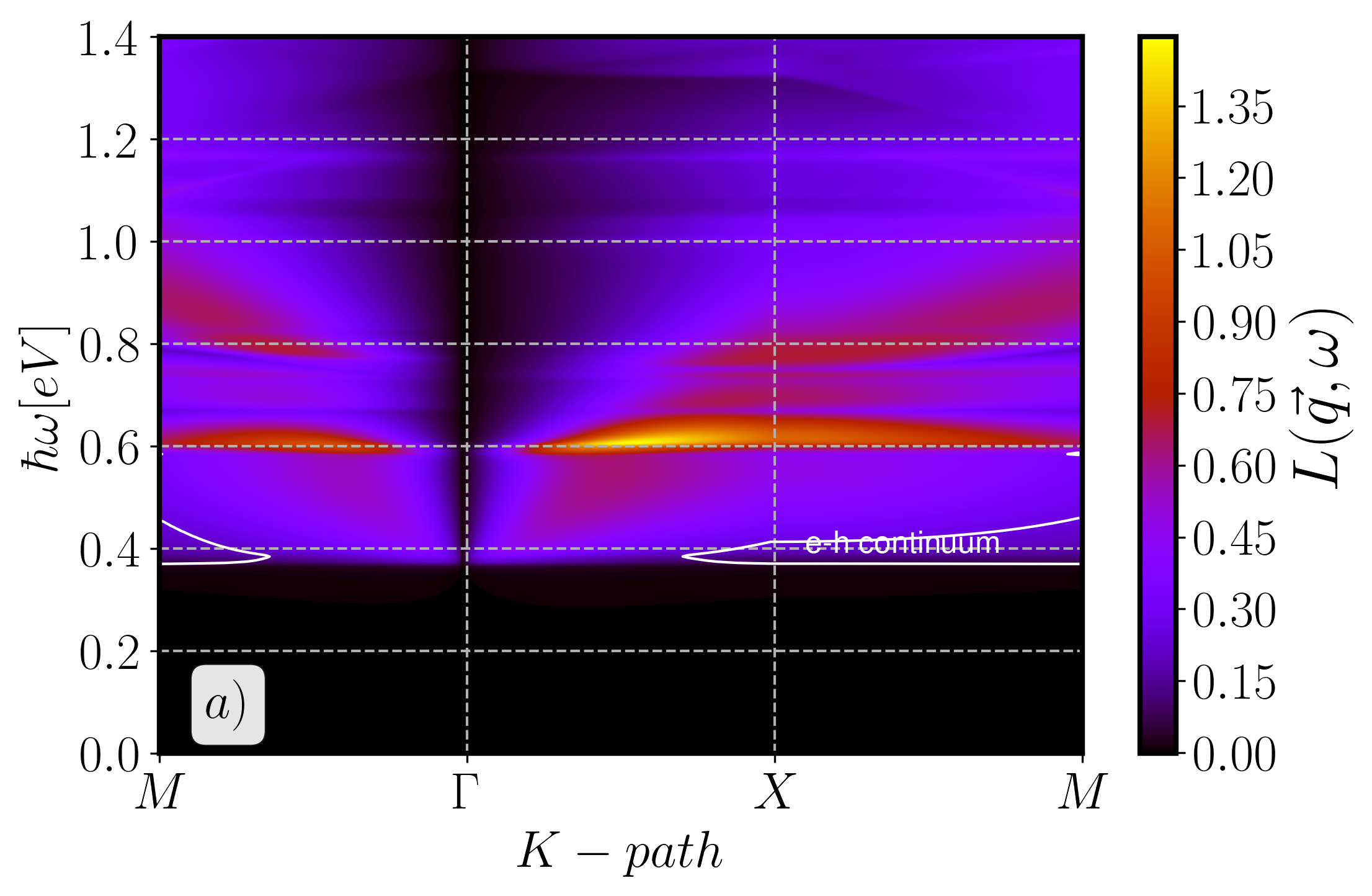}
	\includegraphics[scale=0.5]{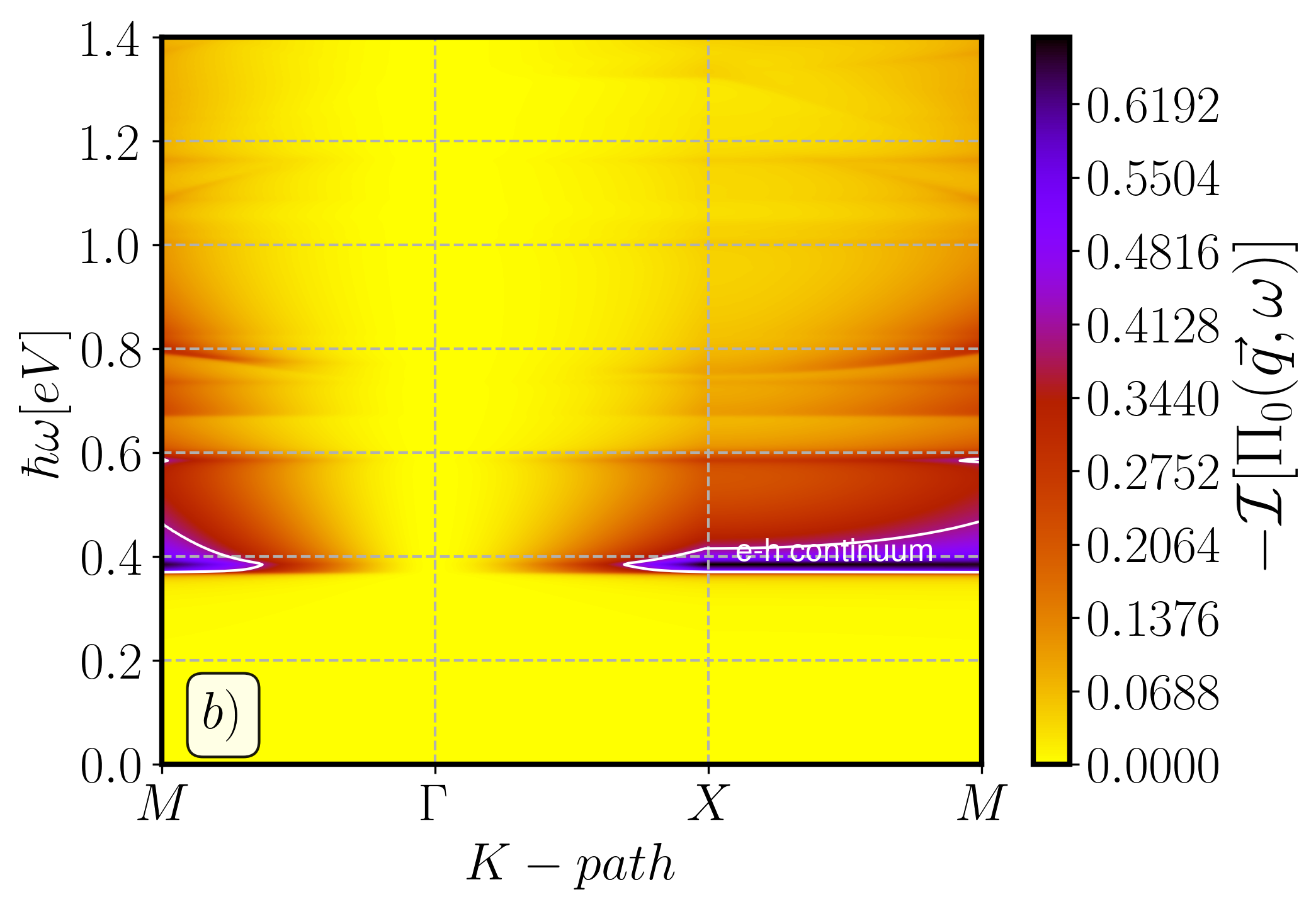}
	\caption{{(a) Plasmon dispersion for $UCHG(7,7,3)$, showing multiple well-defined branches. (b) The corresponding particle-hole continuum, derived from $-\mathcal{I}[\Pi_0(\mathbf{q}, \omega)]$. A significant portion of the plasmon modes, particularly the higher-energy ones, lie outside the damping region (bounded by white lines), indicating they are long-lived excitations.}}
	\label{fig:plasmons-branches-loss-function-773}
\end{figure}

\subsection{Holey graphene 993}

\begin{figure*}[h]
	\centering
	\includegraphics[scale=0.5]{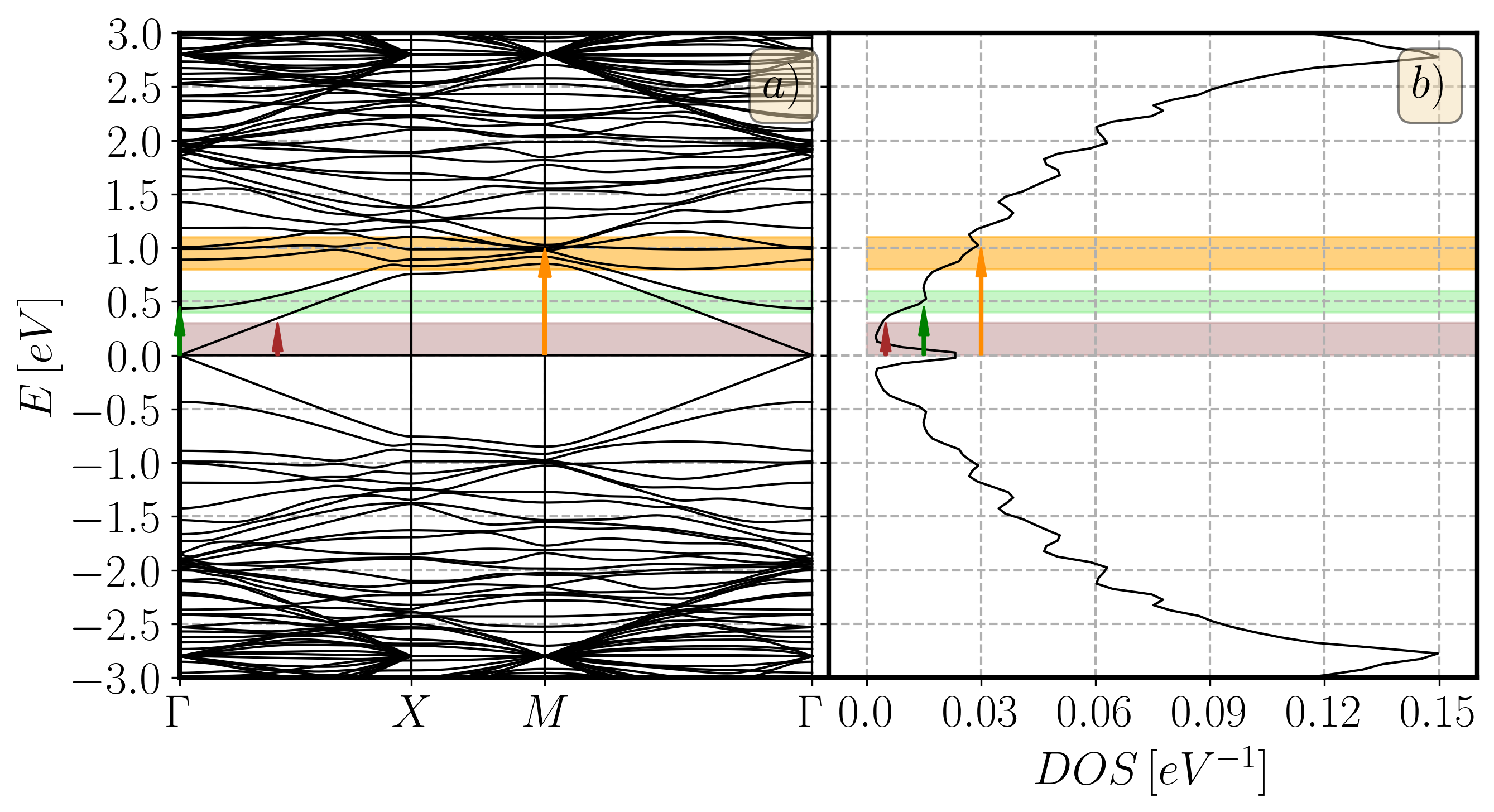}\\
	\includegraphics[scale=0.5]{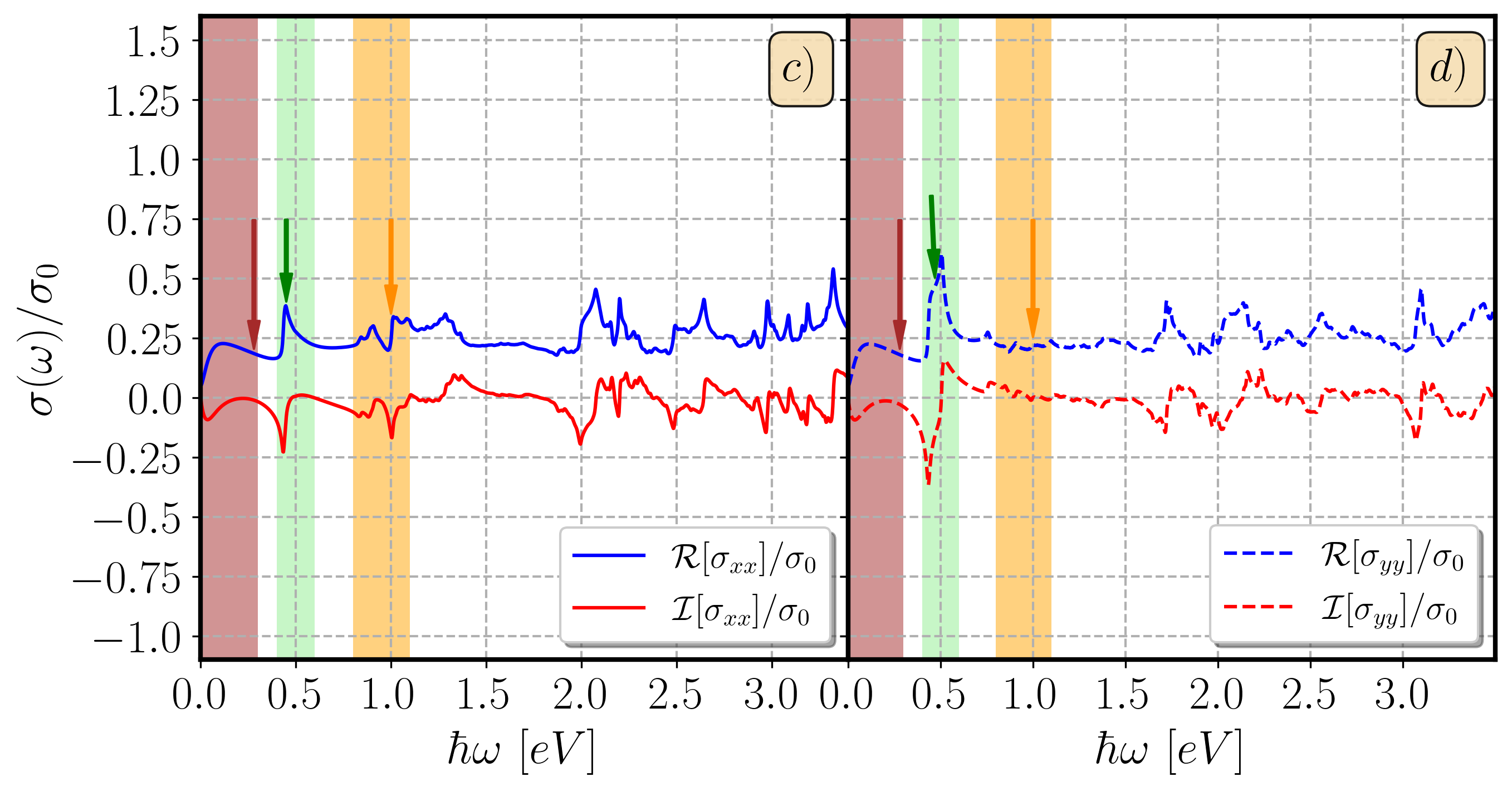}
	\caption{{(a) Electronic band structure and (b) density of states (DOS) for the $UCHG(9,9,3)$ system, which preserves a Dirac cone at the $\Gamma$ point. (c) and (d) Optical conductivities. Unlike the gapped systems, the presence of the Dirac cone results in a non-zero conductivity at low energies. Nevertheless, the system still exhibits significant anisotropy at higher energies (orange region) due to transitions from the flat bands near $X$ and $M$.}}
	\label{fig:bands-dos-holey-993}
\end{figure*}

\begin{figure*}[h]
	\centering
	\includegraphics[scale=0.5]{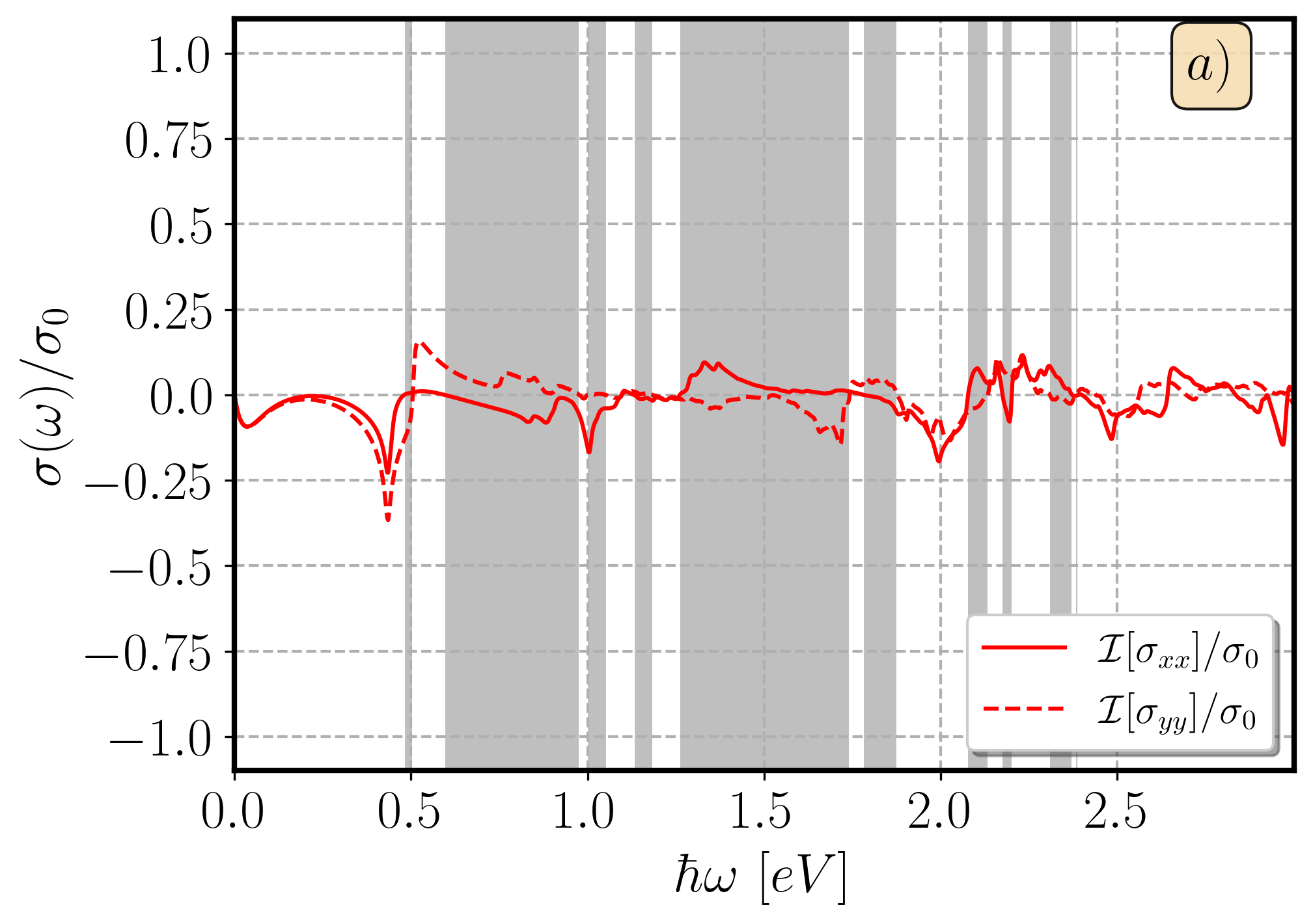}
	\caption{{Imaginary part of the optical conductivity for $UCHG(9,9,3)$. The anisotropy leads to several spectral windows (shaded gray) that support hyperbolic plasmons, demonstrating that this behavior is robust even in gapless HG systems.}}
	\label{fig:hyperbolic-loss-function-993}
\end{figure*}

\begin{figure}[h]
	\centering
	\includegraphics[scale=0.5]{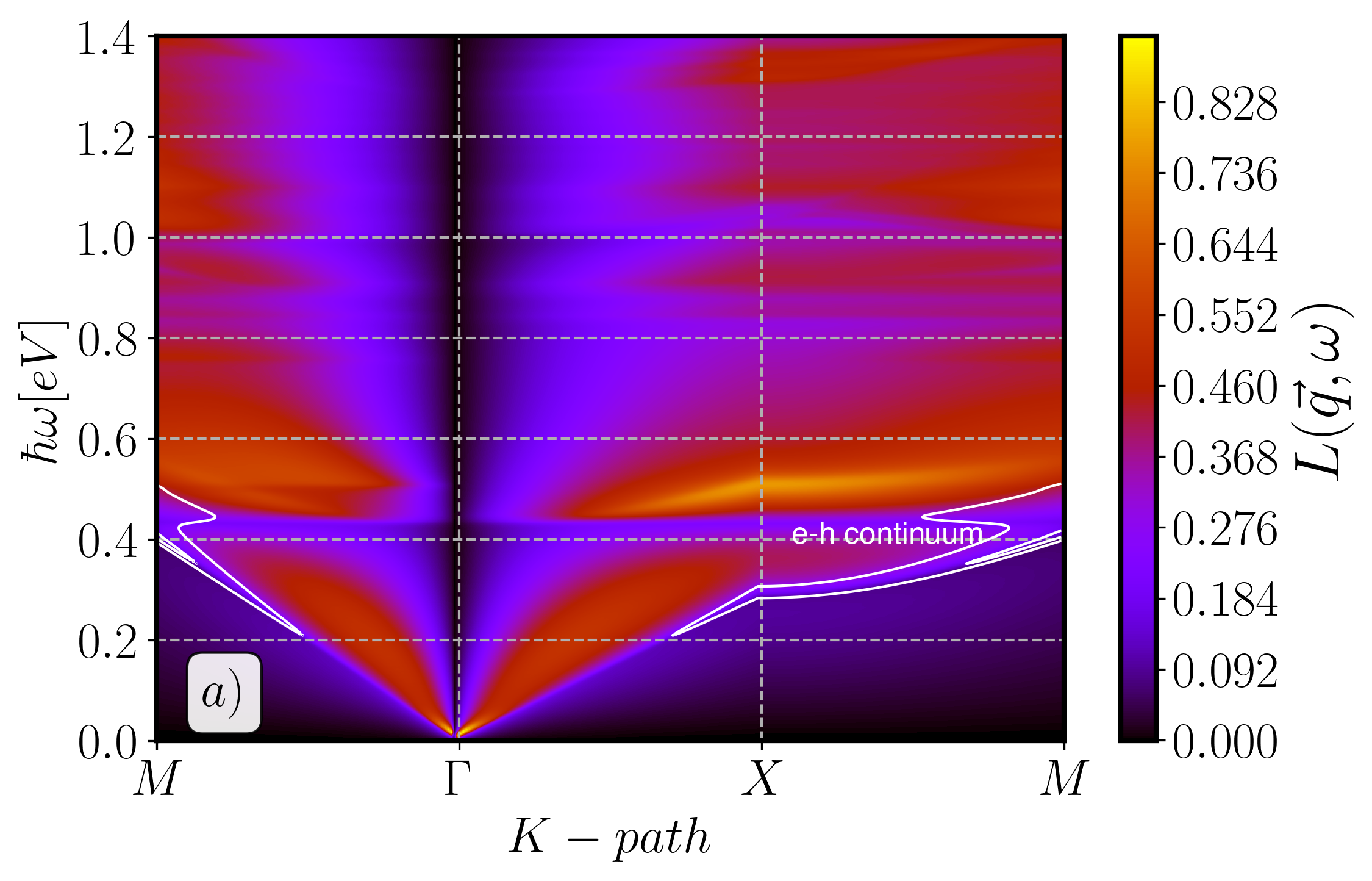}
	\includegraphics[scale=0.5]{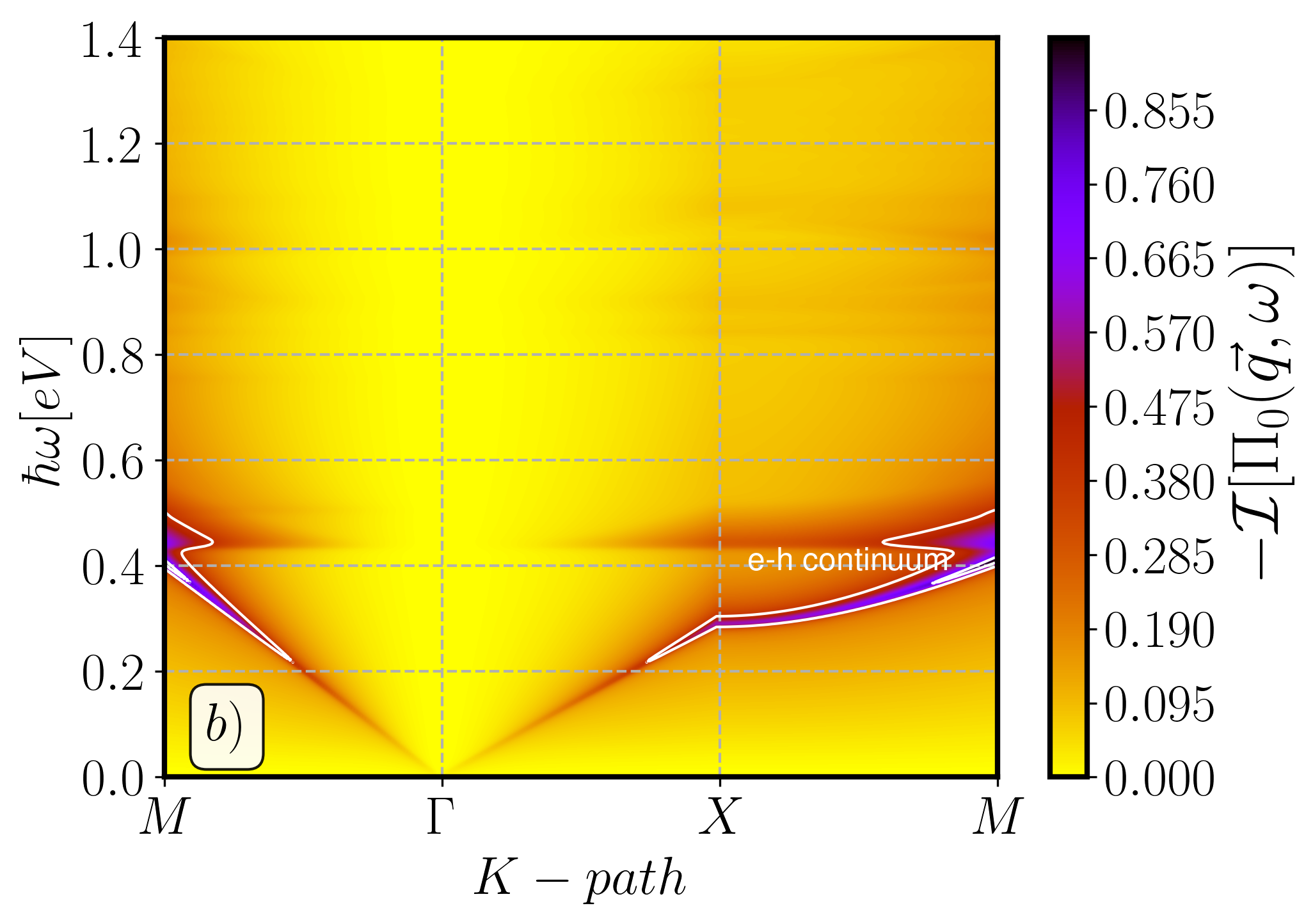}
	\caption{{(a) Plasmon dispersion for the gapless $UCHG(9,9,3)$ system. (b) The particle-hole continuum. The acoustic plasmon mode starting from $\omega=0$ is a characteristic feature of gapless systems. Higher energy optical plasmons are also present and largely undamped.}}
	\label{fig:plasmons-branches-loss-function-993}
\end{figure}

\subsection{Holey graphene 775}

\begin{figure*}[h]
	\centering
	\includegraphics[scale=0.5]{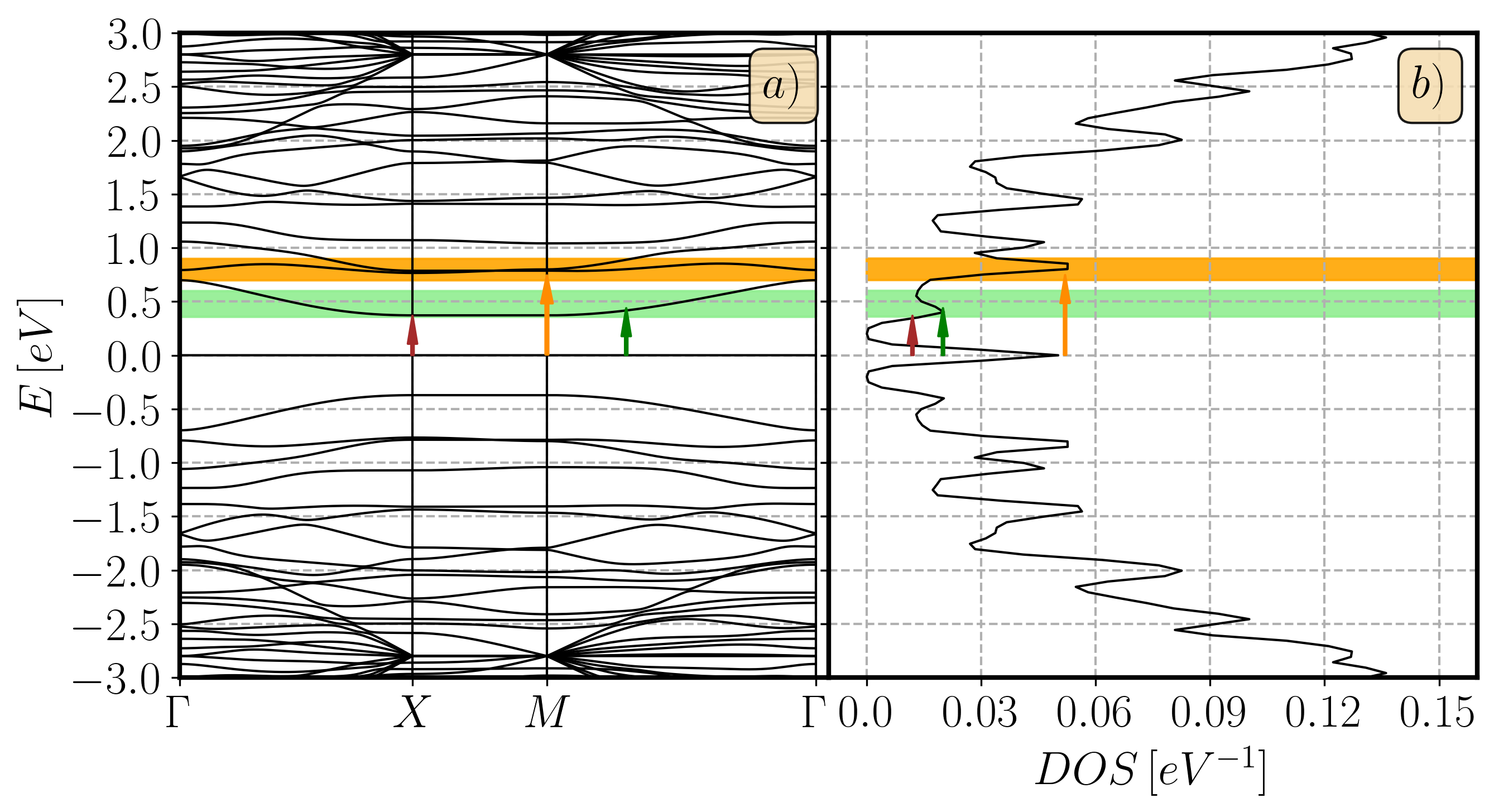}\\
	\includegraphics[scale=0.5]{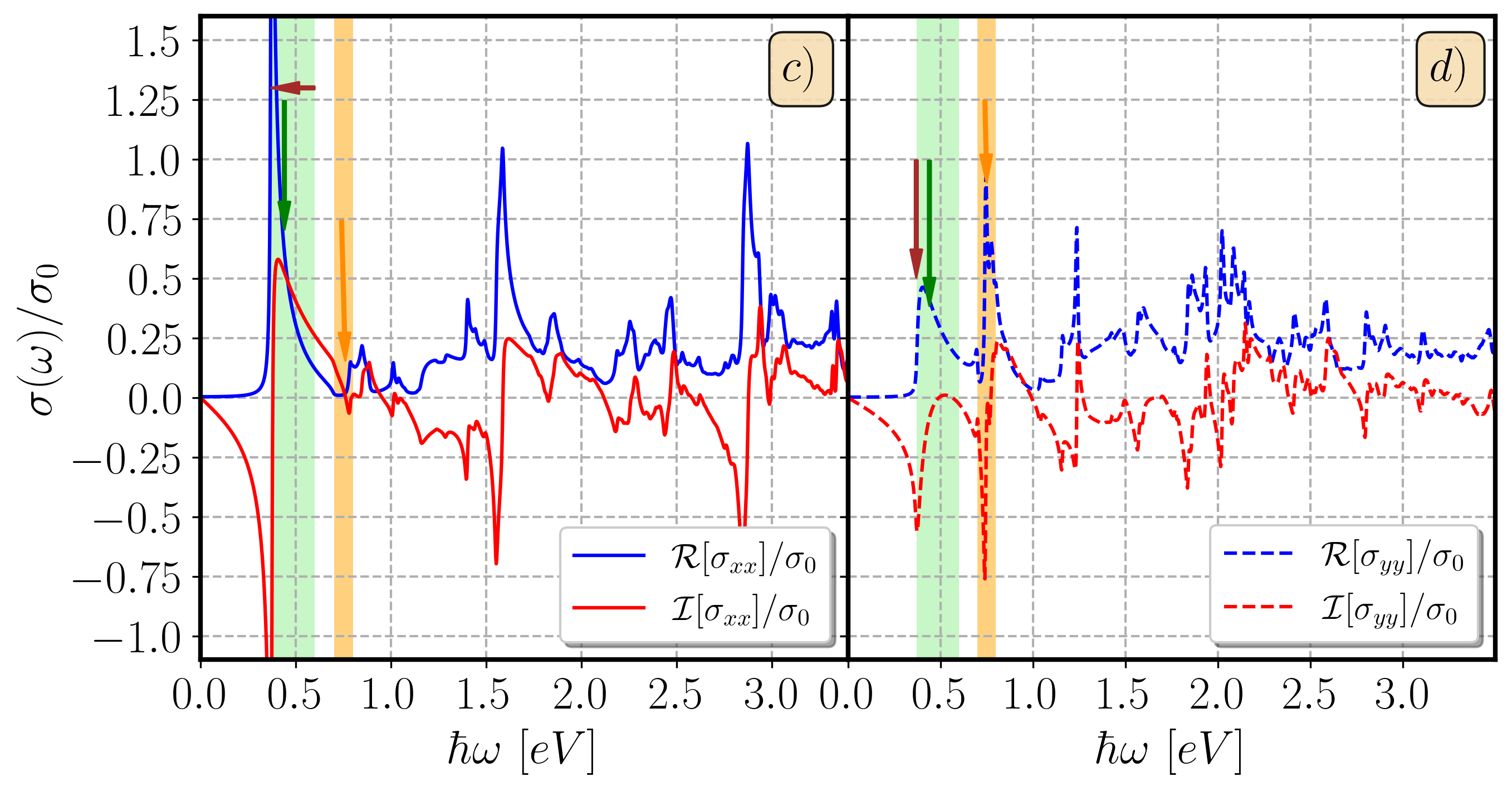}
	\caption{{(a) Band structure and (b) DOS for $UCHG(7,7,5.2)$, which has a larger hole radius compared to UCHG(7,7,3). (c) and (d) The increased hole size enhances the band gaps and flattens the bands further, leading to sharper peaks in the optical conductivity and a more pronounced anisotropy between $\sigma_{xx}$ and $\sigma_{yy}$.}}
	\label{fig:bands-dos-holey-775}
\end{figure*}

\begin{figure*}[h]
	\centering
	\includegraphics[scale=0.5]{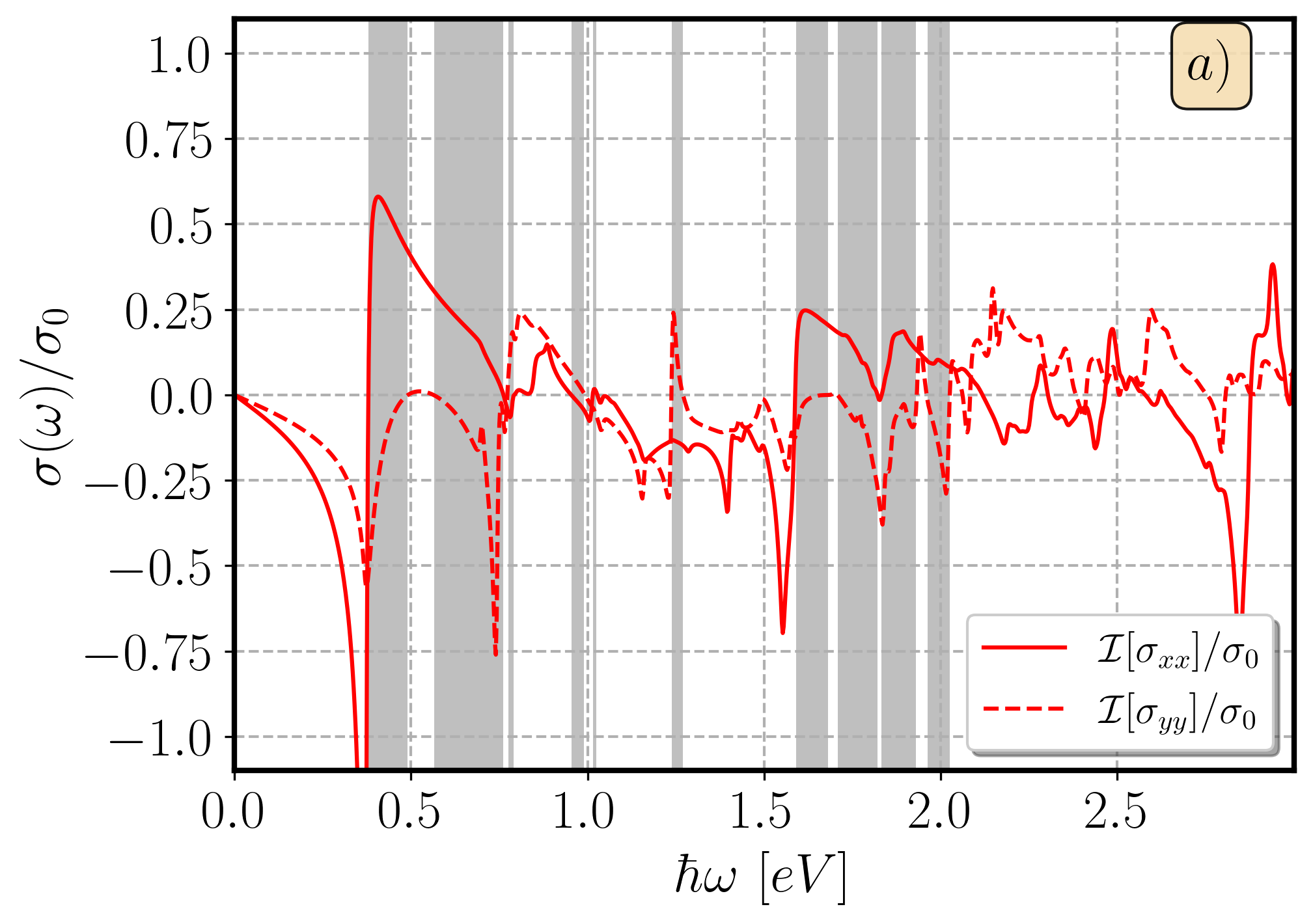}
	\caption{{Imaginary part of the optical conductivity for $UCHG(7,7,5.2)$. The larger hole size results in wider frequency windows for hyperbolic plasmon propagation (shaded gray areas) compared to the smaller-hole counterpart, indicating enhanced anisotropy.}}
	\label{fig:hyperbolic-loss-function-775}
\end{figure*}

\begin{figure}[h]
	\centering
	\includegraphics[scale=0.5]{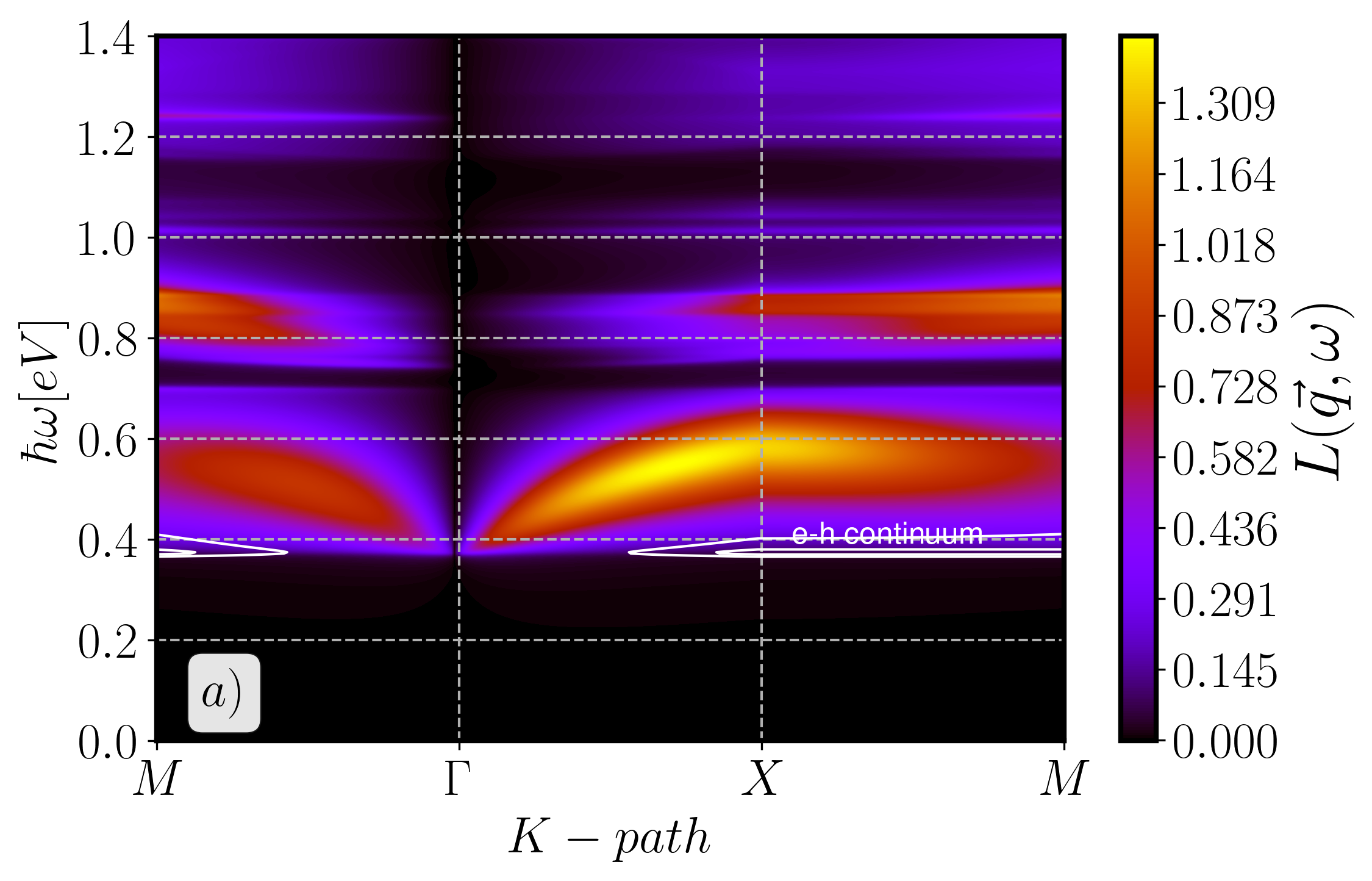}
	\includegraphics[scale=0.5]{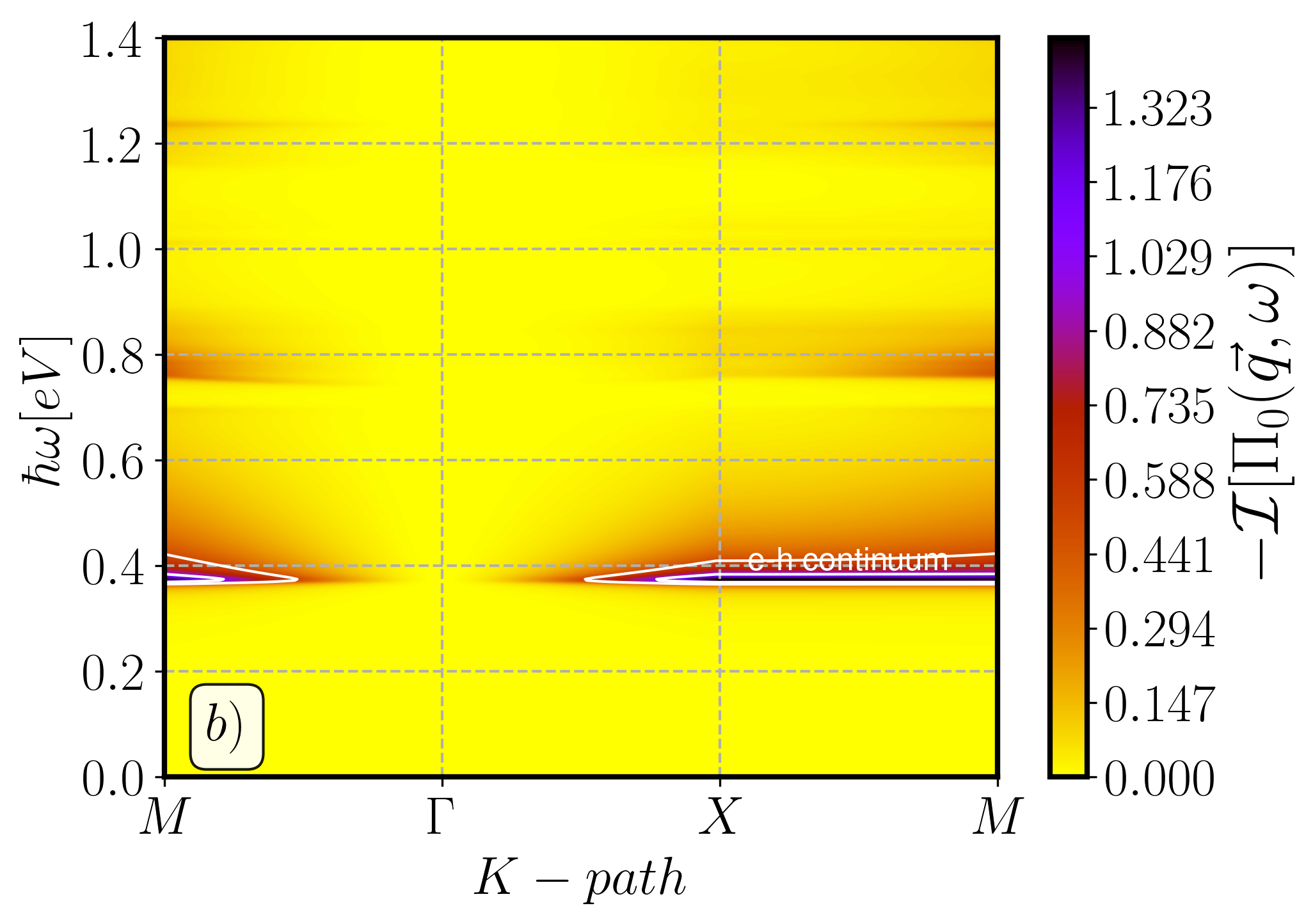}
	\caption{{(a) Plasmon dispersion for $UCHG(7,7,5.2)$. The larger hole size leads to the emergence of even flatter plasmon bands at lower energies compared to the smaller-hole counterparts. (b) The particle-hole continuum. The low-energy flat plasmons are well-defined and exist in a region free of Landau damping.}}
	\label{fig:plasmons-branches-loss-function-775}
\end{figure}

\subsection{Holey graphene 995}

\begin{figure*}[h]
	\centering
	\includegraphics[scale=0.5]{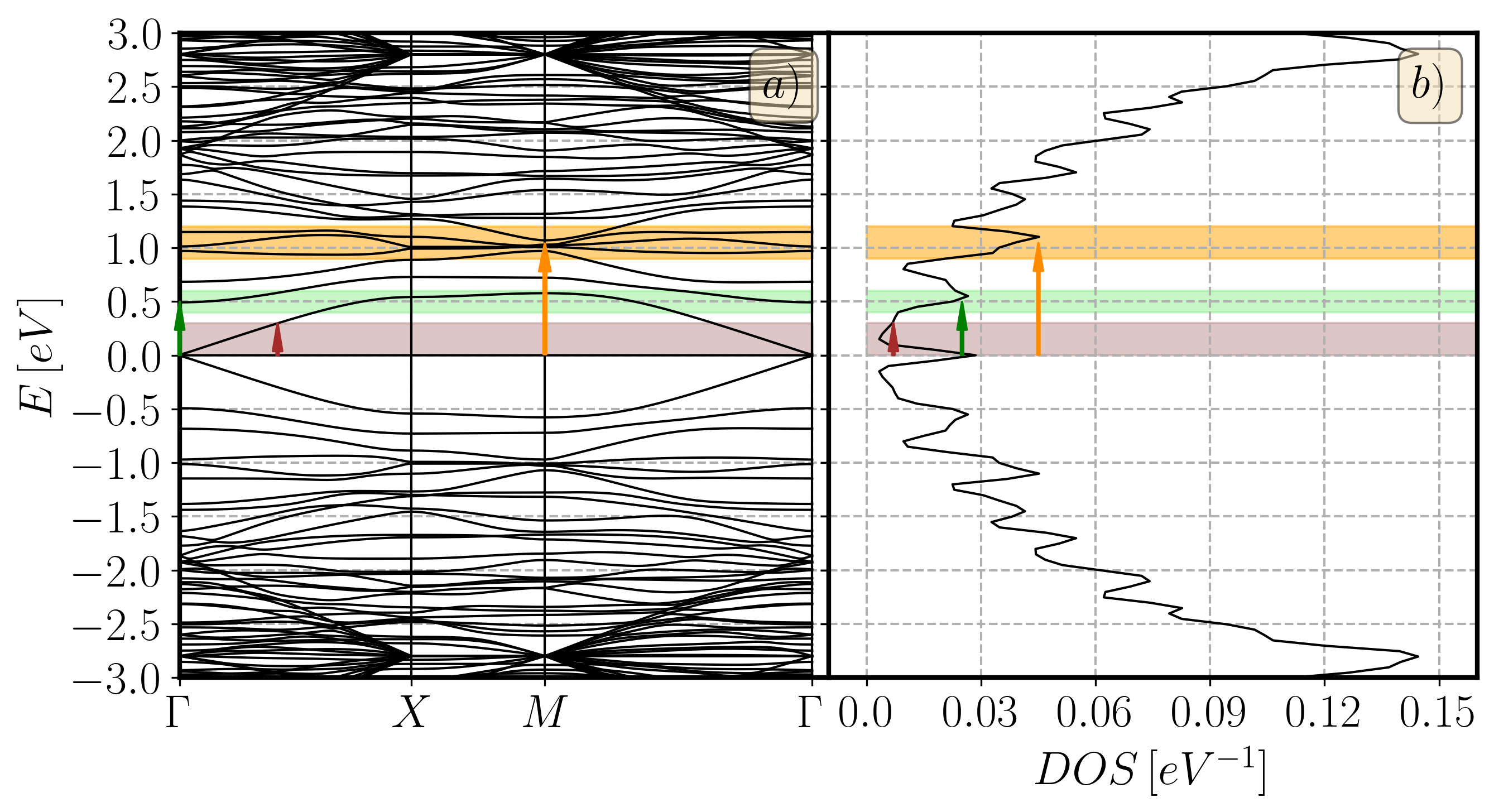}\\
	\includegraphics[scale=0.5]{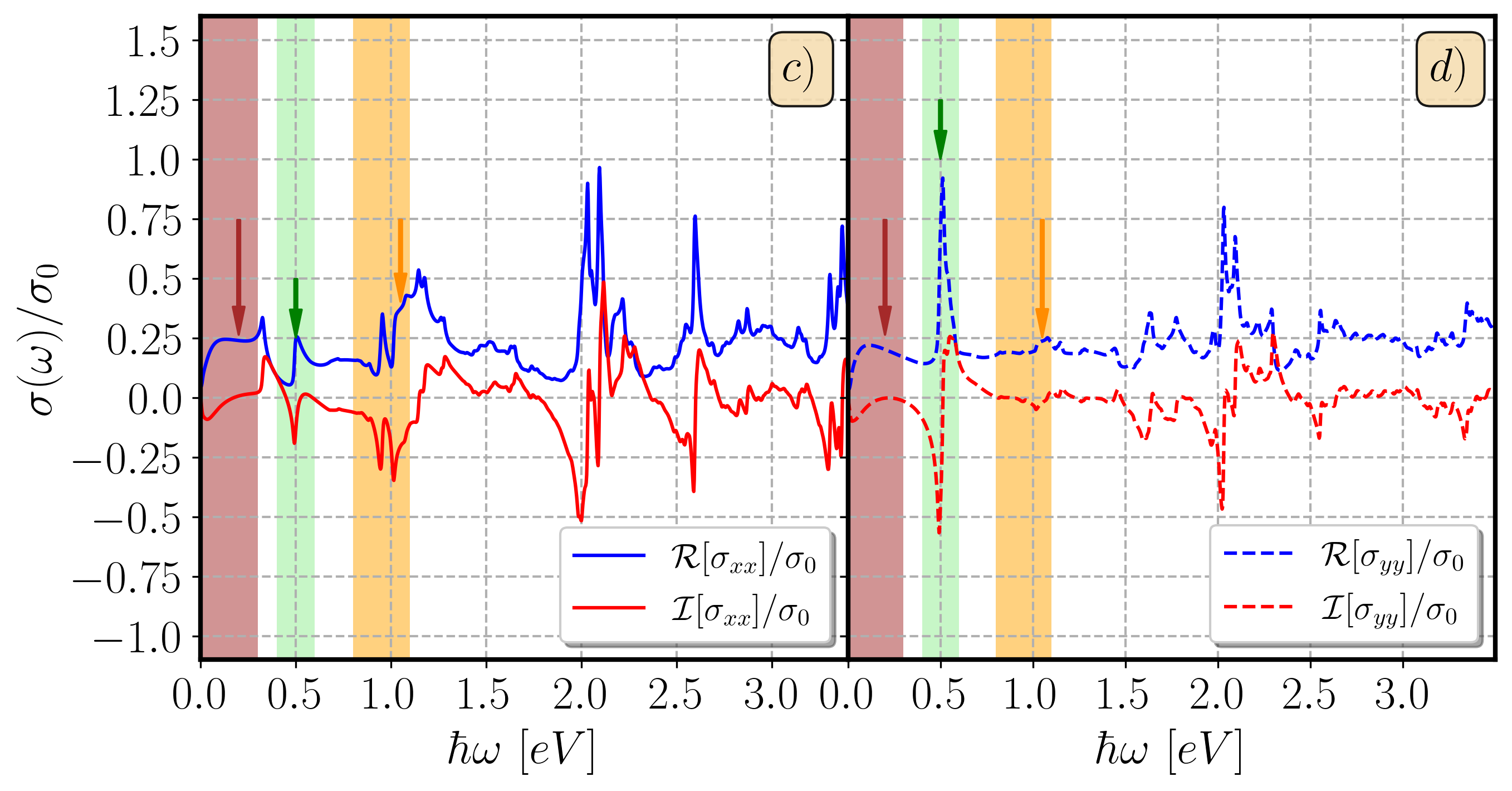}
	\caption{{(a) Band structure and (b) DOS for the gapless $UCHG(9,9,5.2)$ system with a large hole radius. (c) and (d) The optical conductivity shows features of both a gapless system (Drude-like tail) and a highly anisotropic material, with a very strong peak in $\sigma_{yy}$ in the orange-highlighted region.}}
	\label{fig:bands-dos-holey-995}
\end{figure*}

\begin{figure*}[h]
	\centering
	\includegraphics[scale=0.5]{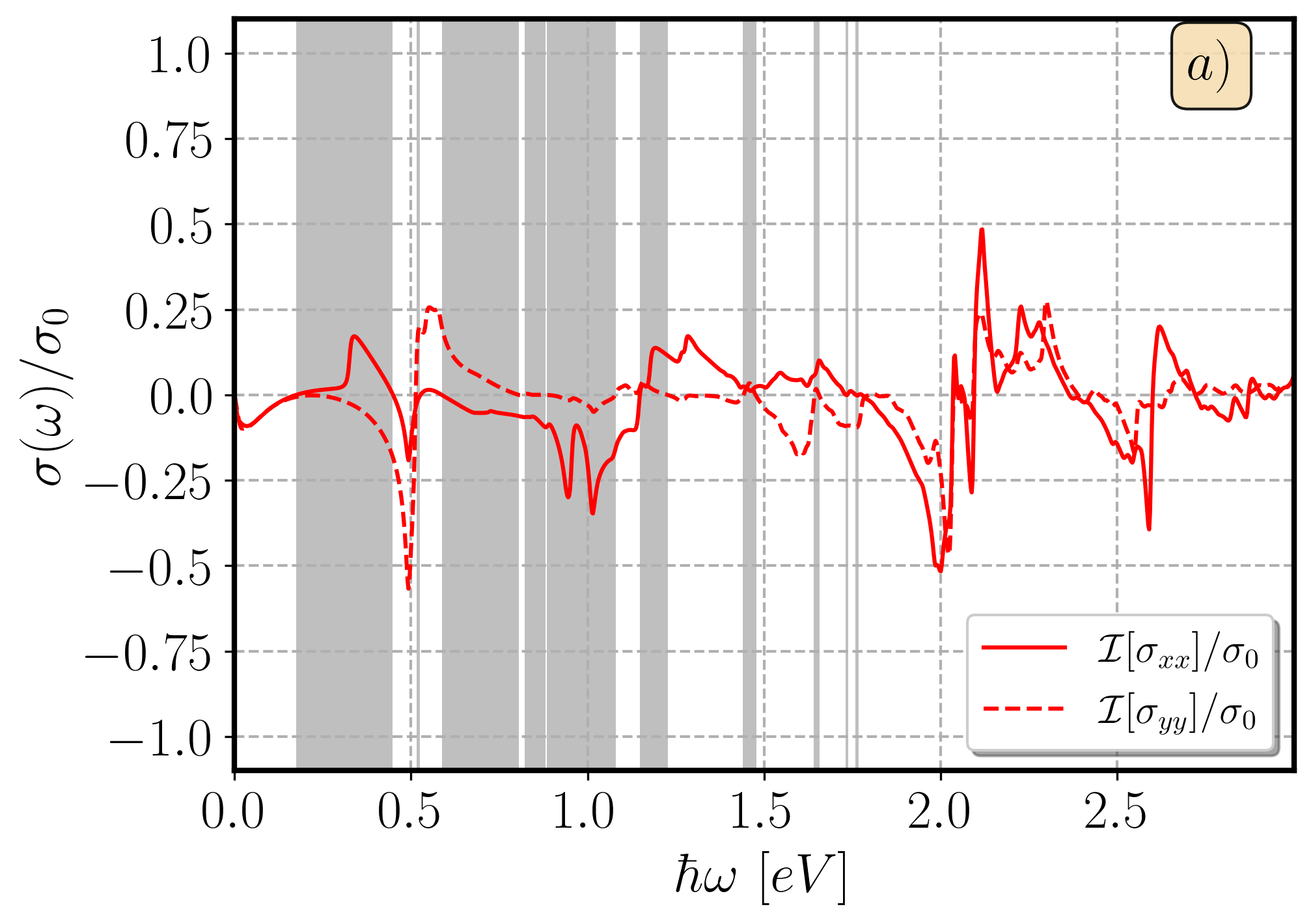}
	\caption{{Imaginary part of the optical conductivity for $UCHG(9,9,5.2)$. This system exhibits broad spectral regions for hyperbolic plasmons, a direct consequence of the strong anisotropy induced by the large periodic holes.}}
	\label{fig:hyperbolic-loss-function-995}
\end{figure*}

\begin{figure}[h]
	\centering
	\includegraphics[scale=0.5]{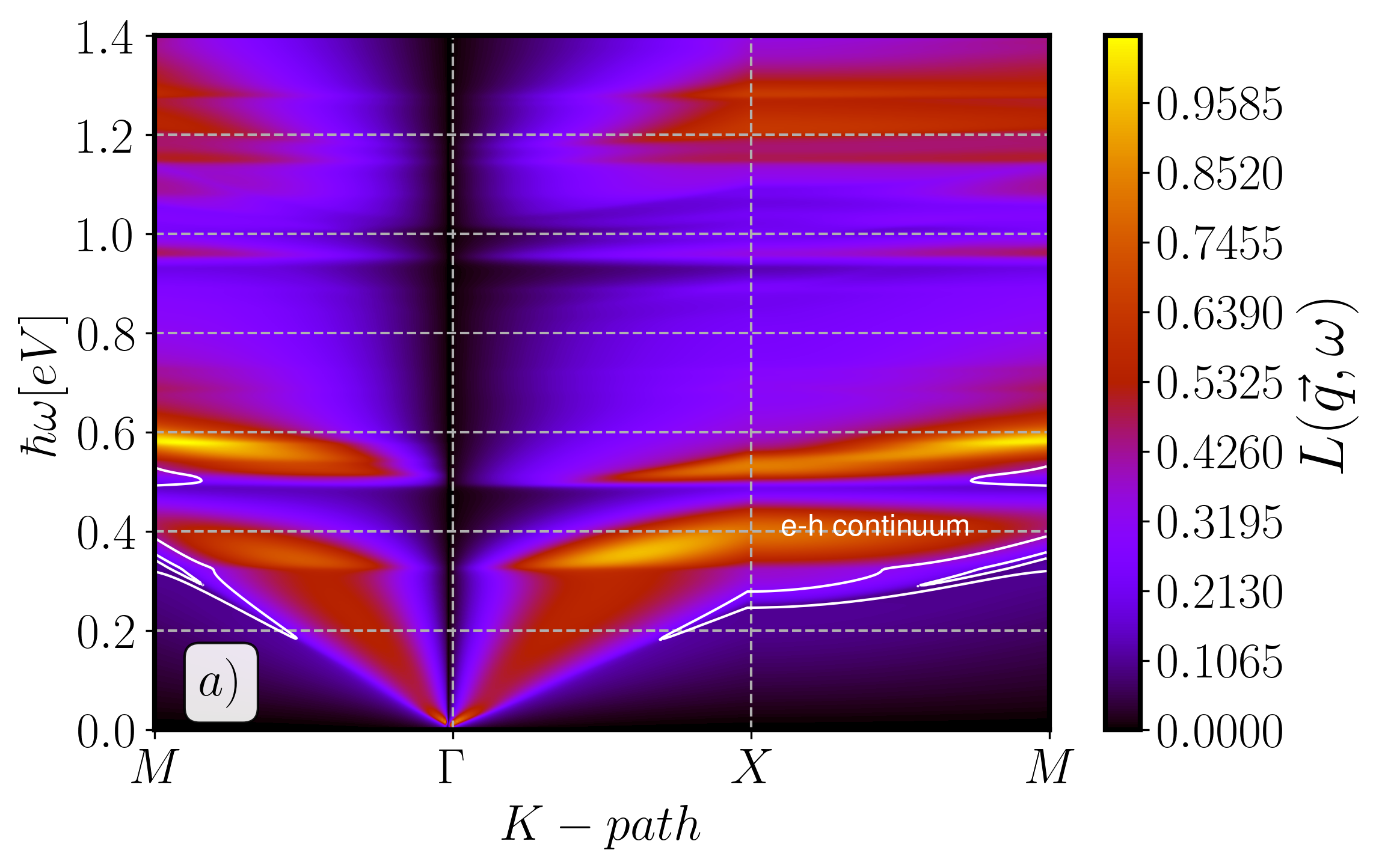}
	\includegraphics[scale=0.5]{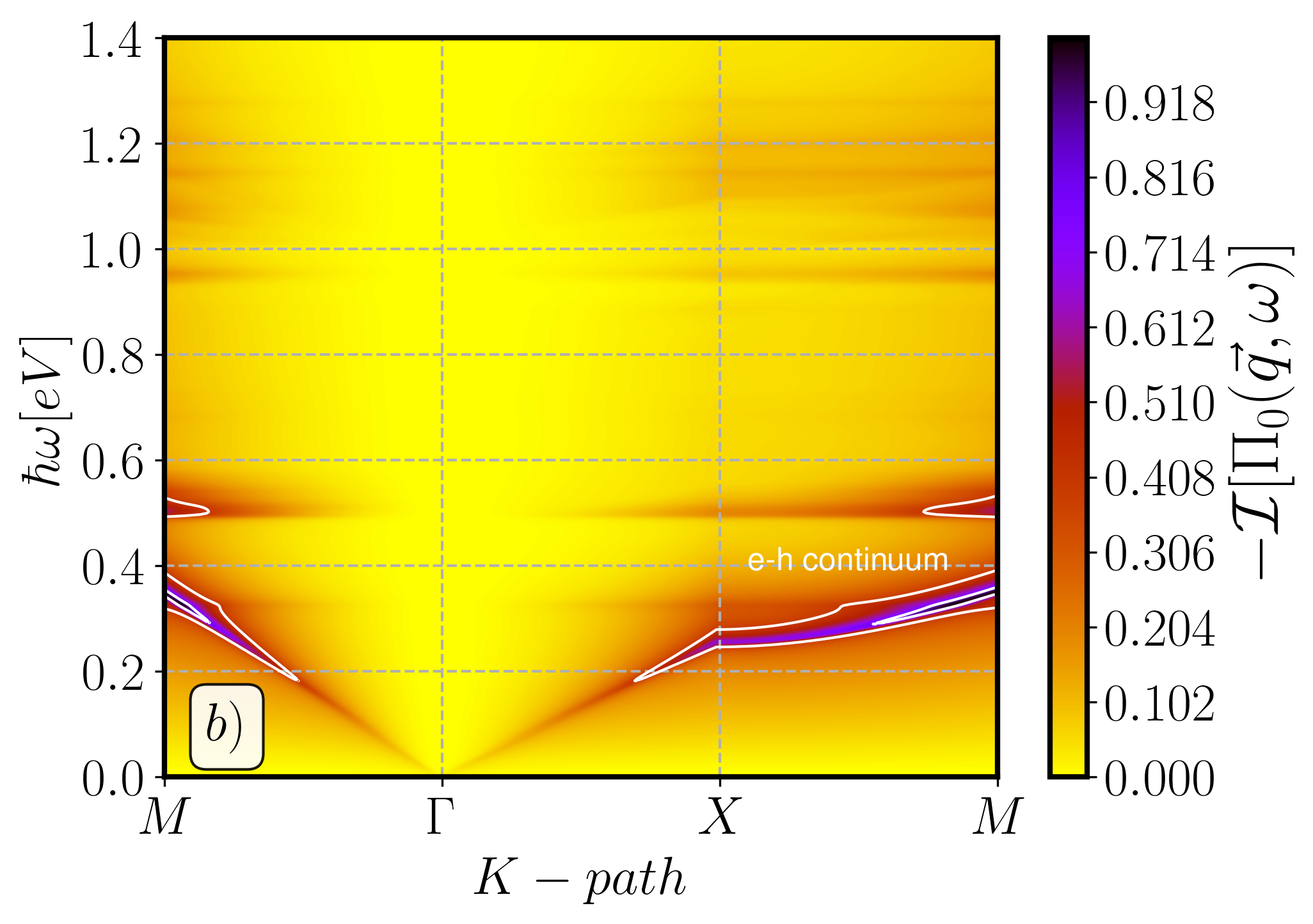}
	\caption{{(a) Plasmon dispersion for $UCHG(9,9,5.2)$. This configuration shows a rich spectrum with multiple flat optical plasmon modes in addition to the acoustic plasmon. (b) The corresponding particle-hole continuum, indicating that these numerous modes are stable against decay into electron-hole pairs.}}
	\label{fig:plasmons-branches-loss-function-995}
\end{figure}

	\section*{References}
	\bibliographystyle{iopart-num}


\end{document}